\documentclass[twocolumn,prb,showpacs,preprintnumbers,amsmath,amssymb,superscriptaddress,bibliography]{revtex4-1}
\usepackage{graphicx}
\usepackage{setspace}

\usepackage{color}
\usepackage{ulem} 
\graphicspath{{./figures/}}
\usepackage{amsmath,amsthm,amssymb}
\usepackage{mathrsfs}

\usepackage[%dvipdfmx,
colorlinks=true, citecolor=blue, urlcolor=blue, linkcolor=blue,
setpagesize=false, bookmarks=false, breaklinks=true
]{hyperref}
\usepackage{lineno}

\newcommand{\heta}{\hat{\eta}}

\newcommand{\hc}{\hat{c}}

\newcommand{\hH}{\hat{H}}

\newcommand{\hn}{\hat{n}}

\newcommand{\eqq}[1]{\begin{align} #1 \end{align}}

\begin{document}
\title{Suppression of Heating by Multi-color Driving Protocols in Floquet Engineered Strongly Correlated Systems}

\author{Yuta Murakami}
\affiliation{Center for Emergent Matter Science, RIKEN, Wako, Saitama 351-0198, Japan}
\author{Michael Sch\"uler}
\affiliation{Laboratory for Theoretical and Computational Physics, Paul Scherrer Institut, 5232 Villigen PSI, Switzerland}
\affiliation{Department of Physics, University of Fribourg, 1700 Fribourg, Switzerland}
\author{Ryotaro Arita}

\affiliation{Center for Emergent Matter Science, RIKEN, Wako, Saitama 351-0198, Japan}
\affiliation{Research Center for Advanced Science and Technology, University of Tokyo, Komaba, Tokyo 153-8904, Japan}
\author{Philipp Werner}
\affiliation{Department of Physics, University of Fribourg, 1700 Fribourg, Switzerland}
\date{\today}

\begin{abstract} 
Heating effects in Floquet engineered system are detrimental to the control of physical properties.
In this work, we show that the heating of periodically driven strongly correlated systems can be suppressed by multi-color driving, i.e., by applying auxiliary excitations which interfere with the absorption processes from the main drive. 
We focus on the Mott insulating single-band Hubbard model and study the effects of multi-color driving with nonequilibrium dynamical mean-field theory.
The main excitation is a periodic electric field with frequency $\Omega$ smaller than the Mott gap, while for the auxiliary excitations, we consider additional electric fields and/or hopping modulations with a higher harmonic of $\Omega$. To suppress the 3-photon absorption of the main excitation, which is a parity-odd process, we consider auxiliary electric-field excitations and a combination of electric-field excitations and hopping modulations. On the other hand, to suppress the 2-photon absorption, which is a parity-even process, we consider hopping modulations. 
The conditions for an efficient suppression of heating are well captured by the Floquet effective Hamiltonian derived with the high-frequency expansion in a rotating frame. 
As an application, we focus on the exchange couplings of the spins (pseudo-spins) in the repulsive (attractive) model, and demonstrate that the suppression of heating allows to realize and clearly observe a significant Floquet-induced change of  the low energy physics.
\end{abstract}

\maketitle

\section{Introduction}
Floquet engineering, where a system is exposed to strong periodic excitations, provides a promising pathway to control physical properties of the system.\cite{Eckardt2017RMP,Oka2019review,Sentef2021review} 
Theoretical predictions of Floquet engineering effects range from the realization of Floquet topological insulators in weakly correlated systems,\cite{Oka2009PRB,Kitagawa2011PRB,Sentef2015,Michael2020PRX} Floquet topological superconductivity,\cite{Takasan2017PRB,Dehghani2021PRR,Kitamura2022} control of band structures~\cite{Dunlap1986PRB,Tsuji2008} and electron-phonon couplings,\cite{Knap2016,Sentef2017,Murakami2017PRB} to control of magnetisms in strongly correlated systems.\cite{Mentink2015,Claassen2017,Kitamura2017PRB,Yonemitsu2021}
Experimentally, driving-induced band renormalizations,\cite{Lignier2007,Eckardt2009,Kilian2019PRL} the creation of topological band structures,\cite{Jotzu2014,Aidelsburger2015} and the control of exchange couplings in strongly correlated systems~\cite{Gorg2018} have been demonstrated in cold atom systems.
In real materials, the realization of a Floquet-induced anomalous Hall effect has also been experimentally reported in graphene,\cite{McIver2020Nature} although the interpretation of the result is not yet fully settled.\cite{Sato2019PRB}
More recently, light-induced anomalous Hall effect are also reported in three-dimensional Dirac systems like Co$_3$SnS$_2$~\cite{Yoshikawa2022} and bismuth~\cite{Hirai2023}.
In addition, giant modifications of the nonlinear response by Floquet engineering have been reported in a correlated material.\cite{Shan2021}
Still, compared to the broad range of theoretical predictions, experimental realizations of Floquet engineering are still limited.

One of the major difficulties in implementing Floquet engineering approaches  in practice is the heating of the system, which may hinder the clear emergence of topological properties and the change in low energy physics.
Heating can occur at different levels. For example, in real systems, there always exist bands above the targeted low-lying bands which most theoretical studies focus on. 
Strong excitations can excite particles from the target bands to the high energy bands, which may prevent the realization of the intended Floquet engineering effect.
Another possibility is excitations within the target bands. Due to the nonthermal or hot distribution caused by such excitations, the control of topological properties and low energy physics becomes difficult to observe.\cite{Michael2020PRX}
A promising way to avoid  heating is to use multi-color driving protocols,\cite{Essliinger2021PRX,Rubio2022PRR} which also provide further controllability of the system.\cite{Neufeld2021PRL,Ikeda2022,Samudra2022PRB,Yixiao2023PRA}
Recently, such a protocol has been implemented in a cold atom system to suppress the excitation of particles to higher energy bands,\cite{Essliinger2021PRX}
while optimal control theory has been used to engineer the band filling in a free system under periodic driving.\cite{Rubio2022PRR}
However, to what extent multi-color driving protocols are beneficial for the Floquet engineering of strongly correlated systems remains to be understood.
Rich low energy physics and phases resulting from the competition or cooperation between various degrees of freedoms are characteristic of strongly correlated systems.
If heating can be suppressed, such low energy physics and phases can be efficiently controlled by Floquet engineering.

In this work, we address this question by analyzing the half-filled single-band Hubbard model, a standard model of strongly correlated systems, with the nonequilibrium dynamical mean-field theory (DMFT).\cite{Georges1996,Aoki2013}
Specifically, we focus on Mott insulators and consider an electric-field excitation with frequency $\Omega$ as the main excitation, which is chosen to be a sub-gap driving.
Additional electric fields and hopping modulations with higher harmonics of $\Omega$ are used as auxiliary excitations, see Fig.~\ref{fig:schematic}(a).
We discuss and demonstrate how the 3-photon and 2-photon absorption can be reduced by suppressing the doublon-holon creation/annihilation terms in the Floquet effective Hamiltonian. The resulting suppression of heating allows us to directly observe the change of the exchange couplings due to the virtual excitations induced by the periodic excitations.

This paper is organized as follows.
In Sec.~\ref{sec:formalism}, we introduce the Hubbard model and derive the corresponding Floquet Hamiltonian using the high-frequency expansion in a rotating frame.
Then, we discuss how the Floquet Hamiltonian allows to determine conditions for the efficient suppression of heating.
In Sec.~\ref{sec:results}, we study the real-time dynamics of the system with a repulsive or attractive interaction using nonequilibrium DMFT, and demonstrate how the suppression of heating in the multi-color protocols works.
Conclusion are given in Sec.~\ref{sec:conclude}.

\section{Formalism} \label{sec:formalism}
\subsection{Model}
In this work, we consider the single-band Hubbard model
\eqq{
\hH(t) = -\sum_{\langle i,j\rangle,\sigma} v_{ij}(t) \hc^\dagger_{i\sigma} \hc_{j\sigma}  + U \sum_j \hn_{j\uparrow} \hn_{j\downarrow}, \label{eq:H_origin}
}
where $\hc^\dagger_{i\sigma}$ is the creation operator for a fermionic particle with spin $\sigma$ at site $i$, $\langle ij\rangle$ indicates a pair of neighboring sites, and 
$\hn_{i\sigma} = \hc^\dagger_{i\sigma} \hc_{i\sigma}$. $U$ is the onsite interaction and $v_{ij}(t)$ is the time-dependent hopping parameter. 
We consider two basic excitation protocols. The first one is an electric-field excitation, which enters the calculation via the Peierls substitution\cite{Aoki2013} $v_{ij}(t) = v_0 e^{i {\bf A}(t)\cdot {\bf r}_{ij}}$.
Here, $v_0$ is the equilibrium hopping parameter, ${\bf A}(t)$ is the vector potential and ${\bf r}_{ij}$ is  the vector from the $j$ site to the $i$ site. The charge of the particle and the bond length are set unity. 
The electric field is related to the vector potential by ${\bf E}(t) = -\partial_t {\bf A}(t)$.
The second protocol is the hopping modulation, which corresponds to $v_{ij}(t) = v_0 + \delta v(t)$.
Both of these protocols have been implemented in cold atom systems~\cite{Eckardt2017RMP, Kilian2019PRL,Essliinger2021PRX}.
In real materials, electric-field excitations can be implemented easily, while hopping modulations may be achieved via the excitation of coherent phonons~\cite{Subedi2014,Mankowsky_2016}.

In this paper, we consider systems on bipartite lattices, i.e. the Bethe lattice or hyper-cubic lattices, at half filling. 
We consider models with strong repulsive or strong attractive interactions to discuss the effects of multi-color driving for different types of orders (low energy physics).
When $U$ is repulsive ($U>0$), the system favors singly occupied sites (singlons) in equilibrium, and exhibits an antiferromagnetic (AF) phase at low enough temperatures.
The absorption of energy of $\mathcal{O}(U)$ creates doubly occupied sites (doublons) and empty sites (holons), which can destroy the low-energy spin order~\cite{Denis2014PRB,Eckstein2016}.
When $U$ is attractive ($U<0$), the system favors doublons and holons in equilibrium, and shows $s$-wave superconductivity, charge order or a coexistence of both at low enough temperatures.  
This physics originates from the SU$(2)$ symmetry which corresponds to the spin SU$(2)$ symmetry of the repulsive Hubbard model via the Shiba transformation.
The absorption of an energy of $\mathcal{O}(|U|)$ in the attractive model creates singlons.

 %%%%%%%%%%%%%%%%%%%%%%%%%%%%%%%%%%%%%%%%%%%%%
 \begin{figure}[tb]
  \centering
    \hspace{-0.cm}
    \vspace{0.0cm}
\includegraphics[width=70mm]{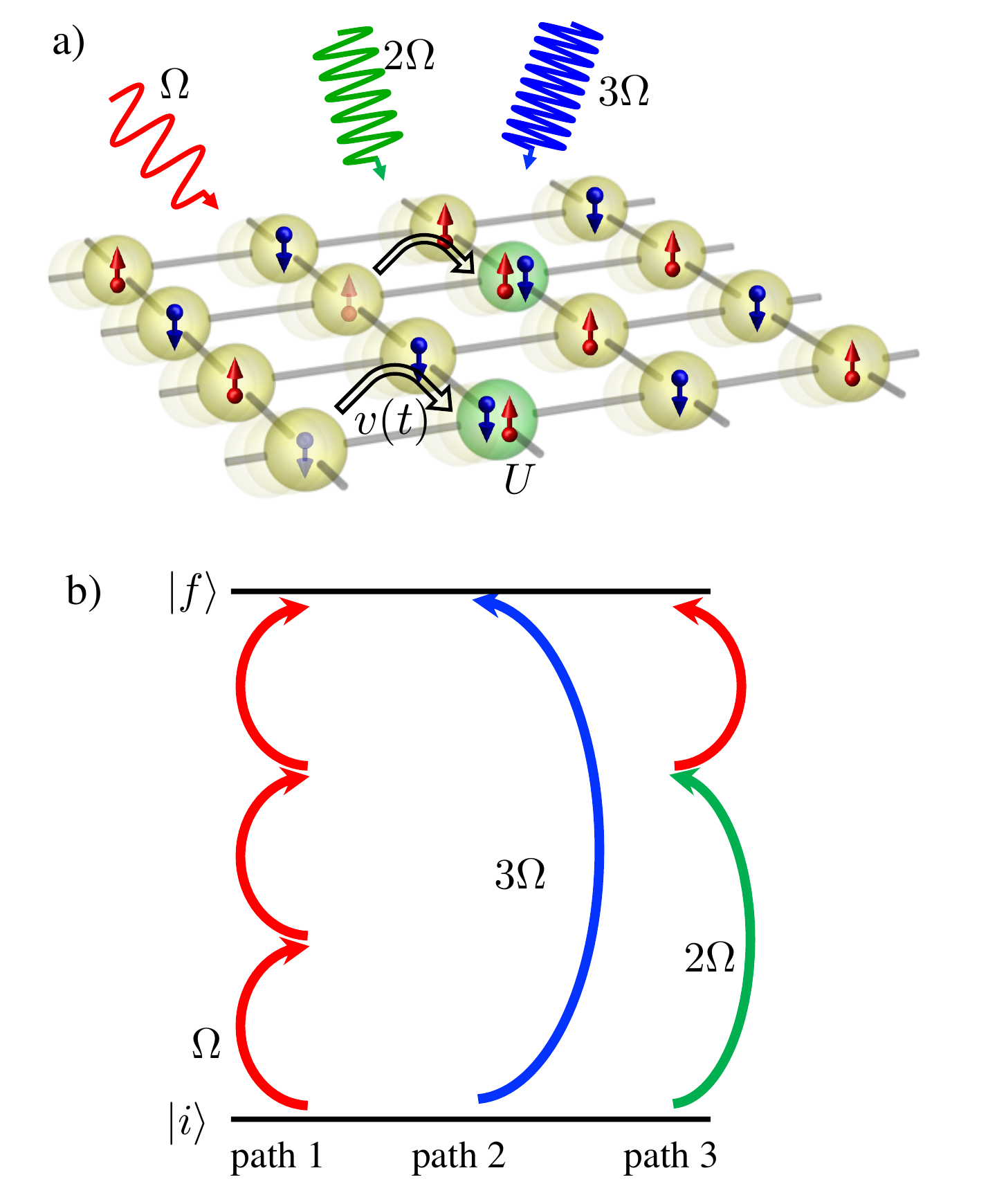} 
  \caption{(a) Schematic picture of the Hubbard model excited with multi-color driving protocols. The red wavy line indicates the main excitation with frequency $\Omega$, while the green and blue wavy lines indicate auxiliary excitations with higher harmonics of $\Omega$. (b) Illustration of the basic idea behind the cancellation of absorption processes using multi-color excitations. The red arrows indicate the $3$ photon excitation from the main drive with frequency $\Omega$.
  The blue and green arrows indicate auxiliary $3\Omega$ and $2\Omega$ excitations, respectively. $|i\rangle$ represents the initial state, while $|f\rangle$ represents the final state.}
  \label{fig:schematic}
\end{figure}
%%%%%%%%%%%%%%%%%%%%%%%%%%%%%%%%%%%%%%%%%%%%

For simplicity, we apply the electric field along a high-symmetry direction.  
In the case of the hypercubic lattices, this is the body-diagonal direction.
More specifically, if we denote the unit vector along the $a$ axis by ${\bf e}_a$, we consider a field along ${\bf e}_{\rm BD} \equiv \sum_a {\bf e}_a$.
As we will mention below, the Bethe lattice can also mimic this situation.
Furthermore, we consider four types of excitation protocols,  see Tab.~\ref{tab:1}. 
The most basic one, which we call ``Type 0", is the conventional single-color electric field excitation. 
This corresponds to $v_{ij}(t) = v_0 e^{i {\bf A}(t)\cdot {\bf r}_{ij}}$ with
\eqq{
{\bf A}(t)= {\bf e}_{\rm BD} A_0 \sin(\Omega t). \label{eq:type0}
}
This electric field excitation with frequency $\Omega$, which is chosen to be smaller than the Mott gap, represents the primary excitation, and we try to suppress the associated heating (absorption) processes with additional weaker excitations.
In the case of ``Type 1" excitations, we consider an additional electric field excitation with frequency $n_1\Omega$, where $n_1$ is an integer.
Specifically, we set $v_{ij}(t) = v_0 e^{i {\bf A}(t)\cdot {\bf r}_{ij}}$ with
\eqq{
{\bf A}(t)= {\bf e}_{\rm BD} (A_0 \sin(\Omega t) + A_1 \sin(n_1\Omega t + \phi_1)). \label{eq:type1}
}
Here $\phi_1$ is the phase shift of the second field relative to the main drive.
In ``Type 2" excitations, we consider an additional hopping modulation with frequency $n_2\Omega$, i.e., 
we set $v_{ij}(t) = v_0(1+\delta v(t)) e^{i {\bf A}(t)\cdot {\bf r}_{ij}}$ with
\eqq{
{\bf A}(t)&= {\bf e}_{\rm BD} A_0 \sin(\Omega t), \nonumber \\
\delta v(t)& = \delta_v \cos(n_2\Omega t + \phi_2). \label{eq:type2}
}
Here $\phi_2$ is the phase shift of the hopping modulation.
In ``Type 3" excitations, we use both an additional electric field excitation with frequency $n_1\Omega$ and a hopping modulation with frequency $n_2\Omega$.
This corresponds to $v_{ij}(t) = v_0(1+\delta v(t)) e^{i {\bf A}(t)\cdot {\bf r}_{ij}}$ with
\eqq{
{\bf A}(t)&= {\bf e}_{\rm BD} (A_0 \sin(\Omega t) + A_1 \sin(n_1\Omega t + \phi_1)),\nonumber \\
\delta v(t)& = \delta_v \cos(n_2\Omega t + \phi_2). \label{eq:type3}
}

 %%%%%%%%%%%%%%%%%%%%%%%%%%%%%%%%%%%%%%%%%%%%%

 \begin{table}[t]
\centering
  \begin{tabular}{|l||c|c|}  \hline
    Name & Involved processes & Relevant Eq.\\ \hline \hline
    Type 0 & single-color electric field & Eq.~\eqref{eq:type0}  \\ \hline
    Type 1 & two-color electric field & Eq.~\eqref{eq:type1}  \\ \hline
    Type 2 & \begin{tabular}{c} single-color electric field +\\ single-color hopping modulation  \end{tabular}& Eq.~\eqref{eq:type2}  \\ \hline
    Type 3 & \begin{tabular}{c} tow-color electric field +\\ single-color hopping modulation  \end{tabular}  & Eq.~\eqref{eq:type3}\\ \hline
  \end{tabular}
    \caption{Summary of excitation protocols. }
    \label{tab:1}
\end{table}
 %%%%%%%%%%%%%%%%%%%%%%%%%%%%%%%%%%%%%%%%%%%%
%%%%%%%%%%%%%%%%%%%%%%%%%%%%%%%%%%%%%%
%%%%%%%%%%%%%%%%%%%%%%%%%%%%%%%%%%%%%%

\subsection{Floquet Hamiltonians in the rotating frame and suppression of absorption}

The idea underlying the suppression of heating with multi-color excitations is based on cancellations between different excitation processes.
Let us consider an excitation process from $|i\rangle$ to $|f\rangle$ using $m$ photons from the electric field excitation of frequency $\Omega$, see Fig.~\ref{fig:schematic}(b).
With additional fields with higher harmonics of $\Omega$, it is possible to create other excitations from $|i\rangle$ to $|f\rangle$, which can interfere with the main excitation process.
In the perturbative regime with respect to the field strength, the amplitude of the $m$-photon process of the main field is $\mathcal{O}(A_0^m)$. 
The amplitude of the interfering processes produced by the additional fields should be of the same order. 
For example, the strength of an additional field $A'$ with frequency $m\Omega$, whose first order processes interfere with the $m$-photon processes of the $\Omega$-frequency field, should be $A'\simeq \mathcal{O}(A_0^m)$.
Hence, $A'$ can be much weaker than the main field.
For this reason, we refer to the additional fields as auxiliary fields.
Although such cancellations can be discussed in detail within the framework of time-dependent perturbation theory, 
an alternative and simpler option is to analyze the Floquet Hamiltonian, as will be done below. 

 %%%%%%%%%%%%%%%%%%%%%%%%%%%%%%%%%%%%%%%%%%%%%
 \begin{figure}[t]
  \centering
    \hspace{-0.cm}
    \vspace{0.0cm}
\includegraphics[width=85mm]{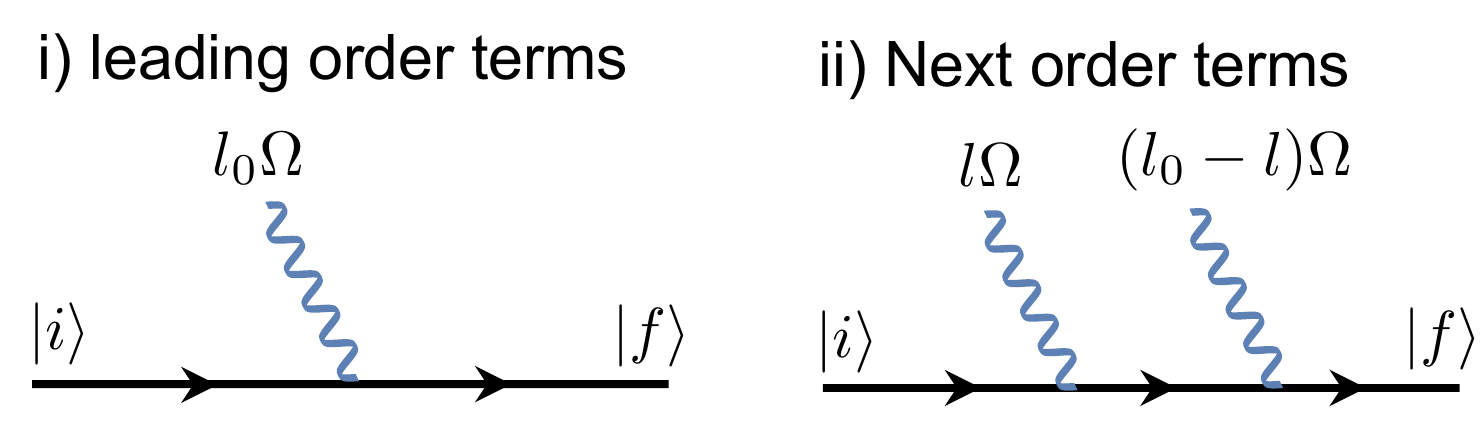} 
  \caption{ Schematic picture of excitation processes 
  corresponding to
  the doublon-holon creation/annihilation terms in the Floquet Hamiltonian. }
  \label{fig:schematic2}
\end{figure}
%%%%%%%%%%%%%%%%%%%%%%%%%%%%%%%%%%%%%%%%%%%%

In the following, we consider subgap excitations where $U=l_0 \Omega + \Delta U$ ($l_0$ is an integer), and assume that $|U|,\Omega\gg |v_0|,|\Delta U|$.
Following Ref.~\onlinecite{Bukov2016PRL}, we can derive the effective Floquet Hamiltonian applying the high frequency expansion in a rotating frame.
This procedure allows to deal with the effects of $U$ and $\Omega$ on equal footing, and the resultant effective Hamiltonian naturally includes the absorption processes and the exchange couplings from virtual excitations due to periodic excitations, as seen below.
First, introducing $U_0 \equiv l_0 \Omega$, we switch to the rotating frame defined by  a unitary transformation $\hat{\mathcal{U}}(t) = \exp (-i U_0 t\sum_j \hn_{j\uparrow} \hn_{j\downarrow})$ by calculating $|\psi^{\rm rot}(t)\rangle = \hat{\mathcal{U}}^\dagger(t) |\psi(t)\rangle$.
The resultant Hamiltonian in the rotating frame ($\hH^{\rm rot}(t) = \hat{\mathcal{U}}^\dagger(t) \hH(t) \hat{\mathcal{U}}(t) + i(\partial_t \hat{\mathcal{U}}^\dagger(t)) \hat{\mathcal{U}}(t)$) is 
\eqq{
\hH^{\rm rot}(t)
& = -\sum_{\langle i,j\rangle,\sigma} \left\{ v_{ij}(t) \hat{g}_{ij\sigma} + [v_{ij}(t) e^{iU_0t} \hat{h}^\dagger_{ij\sigma} + h.c.]  \right\} \nonumber\\
& \;\;\;\;+ \Delta U \sum_j \hn_{j\uparrow} \hn_{j\downarrow}. \label{eq:H_rot}
}
Here we introduced
$\hat{g}_{ij\sigma} = (1-\hn_{i\bar{\sigma}}) \hc^\dagger_{i\sigma} \hc_{j\sigma} (1-\hn_{j\bar{\sigma}}) + \hn_{i\bar{\sigma}} \hc^\dagger_{i\sigma} \hc_{j\sigma} \hn_{j\bar{\sigma}}$ and 
$\hat{h}^\dagger_{ij\sigma} = \hn_{i\bar{\sigma}} \hc^\dagger_{i\sigma} \hc_{j\sigma} (1-\hn_{j\bar{\sigma}})$.
The former operator does not change the number of doublons and holons, and the latter is the generator of them.
Assuming $\Omega\gg |v_0|,|\Delta U|$, we apply the high-frequency expansion to the Hamiltonian \eqref{eq:H_rot}, and obtain the effective Floquet Hamiltonian 
\eqq{
\hH_{\rm eff} = \hH^{\rm rot}_0 + \sum_{l>0} \frac{[\hat{H}_l^{\rm rot},\hat{H}_{-l}^{\rm rot}]}{l\Omega} + \mathcal{O}\Bigl(\frac{1}{\Omega^2}\Bigl).
}
Here $\hH^{\rm rot}(t) = \sum_l \hat{H}_l^{\rm rot} e^{il\Omega t}$.
We note that this effective Hamiltonian can describe the stroboscopic time evolution of the system.~\cite{Bukov_review} 
To express $\hat{H}_l^{\rm rot}$ we introduce the Fourier components $\mathcal{A}^{(l)}$ defined by 
$v_{ij}(t) = v_0 \sum_l \mathcal{A}_{ij}^{(l)}e^{il\Omega t}$
and set $\mathcal{B}_{ij}^{(l)}\equiv \mathcal{A}_{ij}^{(l-l_0)}$, where 
$v_{ij}(t) e^{iU_0t}= v_0 \sum_l \mathcal{A}_{ij}^{(l)}e^{i(l+l_0)\Omega t}= v_0 \sum_l \mathcal{B}_{ij}^{(l)}e^{i l\Omega t}$ is full-filled.

The lowest order Hamiltonian is 
\eqq{
\hH^{\rm rot}_0 =& -v_0 \sum_{\langle ij\rangle \sigma} \left\{ \mathcal{A}^{(0)}_{ij} \hat{g}_{ij\sigma}  + [\mathcal{B}^{(0)}_{ij} \hat{h}^\dagger_{ij\sigma}  + h.c. ]\right\} \nonumber \\
& + \Delta U \sum_j \hn_{j\uparrow} \hn_{j\downarrow}.\label{eq:Heff1_0}
}
Remember that $\mathcal{B}^{(0)}_{ij} =  \mathcal{A}_{ij}^{(-l_0)}$ is a Fourier component of the time-dependent hopping amplitude of the original Hamiltonian \eqref{eq:H_origin},
and corresponds to the amplitude for the instantaneous absorption of the energy $l_0\Omega$.
For $l\neq 0$, we have 
\eqq{
\hH^{\rm rot}_l = -v_0 \sum_{\langle ij\rangle \sigma} \left\{ \mathcal{A}^{(l)}_{ij} \hat{g}_{ij\sigma}  + [\mathcal{B}^{(l)}_{ij} \hat{h}^\dagger_{ij\sigma}  + \mathcal{B}_{ij}^{(-l)*}\hat{h}_{ij\sigma}] \right\}. \label{eq:Heff1}
}
Therefore, as far as the leading order Hamiltonian is concerned, if $\mathcal{B}^{(0)}=0$, there is no creation of doublons and holons. 

The second order term $\mathcal{O}(\frac{v^2_0}{\Omega})$ can be expressed as $\hH_\text{eff}^{(2)}=\hH_\text{eff,1}^{(2)} + \hH_\text{eff,2}^{(2)} + \hH_\text{eff,3}^{(2)}$
with 
\begin{subequations}\label{eq:Heff2}
\eqq{
&\hH_\text{eff,1}^{(2)} = \sum_{l>0} \frac{v_0^2}{l\Omega} \Bigg[ \sum_{\langle ij\rangle \sigma} \mathcal{A}^{(l)}_{ij} \hat{g}_{ij\sigma},\sum_{\langle i'j'\rangle \sigma'} \mathcal{A}^{(-l)}_{i'j'} \hat{g}_{i'j'\sigma'} \Bigg], \\
&\hH_\text{eff,2}^{(2)} \label{eq:H_eff_2}  \nonumber\\
&= \sum_{l\neq 0} \frac{v_0^2}{l\Omega} \Bigg[ \sum_{\langle ij\rangle \sigma} \mathcal{A}^{(l)}_{ij} \hat{g}_{ij\sigma},\sum_{\langle i'j'\rangle \sigma'} \!\!\!\big(\mathcal{B}^{(-l)}_{i'j'} \hat{h}^\dagger_{i'j'\sigma'}  + \mathcal{B}_{i'j'}^{(l)*}\hat{h}_{i'j'\sigma'} \big) \Bigg],\\
&\hH_\text{eff,3}^{(2)}  = \sum_{l\neq 0} \frac{v_0^2}{l\Omega} \Bigg[ \sum_{\langle ij\rangle \sigma} \mathcal{B}^{(l)}_{ij} \hat{h}^\dagger_{ij\sigma}  ,\sum_{\langle i'j'\rangle \sigma'} \mathcal{B}_{i'j'}^{(l)*}\hat{h}_{i'j'\sigma'} \Bigg].
}
\end{subequations}
$\hH_\text{eff,2}^{(2)}$ describes the creation and annihilation processes of doublons and holons, 
and is relevant for absorption (heating) processes.\cite{Herrmann_2017}
The remaining terms do not change the number of doublons and holons and govern the low energy physics, which we discuss in detail in the next section.
The full expressions for $\hH_\text{eff,2}^{(2)}$ and $\hH_\text{eff,3}^{(2)}$ are given in Appendix~\ref{sec:H_eff}. 
Note that the expressions ~\eqref{eq:Heff1_0}, \eqref{eq:Heff1} and \eqref{eq:Heff2} are generic and not limited to electric-field excitations along ${\bf e}_{\rm BD}$.

Now let us discuss the meaning of the doublon-holon (d-h) creation/annihilation terms in the effective Floquet Hamiltonian, i.e. Eqs.~\eqref{eq:Heff1_0} and \eqref{eq:Heff2}.
Firstly, these terms only describe the $l_0 \Omega$ absorption processes, while possible $l(\neq l_0)\Omega$ absorption processes are not included. 
To express the latter processes, one needs to consider the effective Hamiltonian in the rotating frame of  $e^{il\Omega}$.
Secondly,  $\mathcal{A}^{(l)}$ corresponds to simultaneous $l\Omega$ excitations, as can be seen from its definition.
Therefore, the leading order d-h creation/annihilation processes in Eq.~\eqref{eq:Heff1_0}
correspond to the simultaneous absorption of $l_0 \Omega$,
while the next-leading order terms in Eq.~\ref{eq:H_eff_2}
correspond to the absorption of  $l\Omega$ and $(l_0-l)\Omega$ at different times, see Fig.~\ref{fig:schematic2}.
Thirdly, $\mathcal{A}^{(l)}$ already includes the contribution from different excitation processes,  since $v(t)$ depends in a nonlinear manner on $A(t)$, 
see Eqs. \eqref{eq:demo1} and \eqref{eq:demo2} below for example. 
This allows to suppress some of $\mathcal{A}^{(l)}$ by tuning the parameters of the auxiliary fields. 
We also emphasize that the effective Hamiltonian can be applied in the non-perturbative regime with respect to the field strength.

We note that the effects of the interference between the main field and the auxiliary fields can appear in the effective Floquet Hamiltonian at different levels.
On the one hand, it can affect the value of $\mathcal{A}^{(l)}$ and thus can modify the coefficients of the doublon-holon creation/annihilation terms.
On the other hand, interferences can also occur between different processes expressed by different terms in the effective Hamiltonian.
In the following, to find the conditions for the efficient suppression of heating, we follow the strategy to suppress the d-h creation/annihilation terms in the Floquet Hamiltonian order by order.
This strategy should work in the high-frequency limit ($\Omega\gg |v_0|,|\Delta U|$), but for moderate values of $\Omega$ it is not a priori clear how well this strategy works.
For example, a condition which eliminates the leading order terms may enhance contributions from the next order.
Furthermore, this strategy does not take into account the potential interferences between the different processes described by the terms representing different orders.
It aims to suppress the $l_0 \Omega$ absorption processes only, which is reasonable when such processes are the dominant ones.
We also note that the auxiliary field may enhance or suppress $l\Omega$ ($l\neq l_0$) absorptions.
Despite these potential difficulties, we will show that the Floquet Hamiltonian serves as a useful guide to determine conditions for the efficient suppression of heating in the present system.

For an electric-field excitation along   ${\bf e}_{\rm BD}$, $\mathcal{A}_{ij}^{(l)}$ can take only two possible values for a given $l$, namely $\mathcal{A}_{{\bf e}_a}^{(l)}$ or $\mathcal{A}_{-{\bf e}_a}^{(l)}$. 
In the following, we denote them by $\mathcal{A}_{\bf e}^{(l)}$ and $\mathcal{A}_{-{\bf e}}^{(l)}$, respectively. The same notation is used for $\mathcal{B}_{ij}^{(l)}$.
With this, the coefficients appearing in $\hH_\text{eff,2}^{(2)}$ can be classified into the following four cases and their complex conjugates:
\eqq{
I^{\rm (dh)}_1 = \sum_{l\neq0} \frac{1}{l} \mathcal{A}^{(l)}_{\bf e} \mathcal{B}^{(-l)}_{\bf e}, \;\; I^{\rm (dh)'}_1 = \sum_{l\neq0} \frac{1}{l} \mathcal{A}^{(l)}_{-\bf e} \mathcal{B}^{(-l)}_{-\bf e}, \nonumber \\
I^{\rm (dh)}_2 = \sum_{l\neq0} \frac{1}{l} \mathcal{A}^{(l)}_{\bf e} \mathcal{B}^{(-l)}_{-\bf e}, \;\; I^{\rm (dh)'}_2 = \sum_{l\neq0} \frac{1}{l} \mathcal{A}^{(l)}_{-\bf e} \mathcal{B}^{(-l)}_{\bf e}. 
}
The relevant question then becomes how to suppress $\mathcal{B}^{(0)}$ and these coefficients.

%%%%%%%%%%%%%%%%%%%%%%%%%%%%%%%%%%
%%%%%%%%%%%%%%%%%%%%%%%%%%%%%%%%%%

\subsection{Floquet-Engineered exchange couplings}
From $\hH_\text{eff,1}^{(2)} + \hH_\text{eff,3}^{(2)}$, one can obtain the effective model for the low energy physics.
In particular, $\hH_\text{eff,3}^{(2)}$ includes the exchange terms for spins and pseudo-spins (doublons and holons), 
which control the low-energy physics if $\mathcal{O}(|U|)$ absorption from the equilibrium state is absent.
These terms yield Heisenberg-type Hamiltonians both in the repulsive and attractive cases. 
To simplify the expression of the Hamiltonian, we focus on the case where $\mathcal{A}_{\bf e}^{(l)}=\mathcal{A}_{-\bf e}^{(l)}(-)^l$ and these coefficients are real.
This condition is fulfilled for the four excitation protocols mentioned above if $\phi_1=\phi_2=0$, $n_1={\rm odd}$ and $n_2={\rm even}$.
We mainly focus on this situation in the following.
We note that this condition implies  $I^{\rm (dh)'}_1 \propto I^{\rm (dh)}_1 $, $I^{\rm (dh)'}_2 \propto I^{\rm (dh)}_2$ and $\hH_\text{eff,1}^{(2)}=0$.

When $U$ is repulsive, the low energy physics is described by the spin degrees of freedom.
From  $\hH_\text{eff,3}^{(2)}$ (see Appendix~\ref{sec:H_eff}), the low energy Hamiltonian consisting of the spin exchange terms  can be expressed as 
\eqq{
 \hH_{\rm spin} = J_{\rm s}^{\rm (HE)} \sum_{(ij)} \hat{\bf s}_i \cdot \hat{\bf s}_j. \label{eq:H_spin}
}
Here $(ij)$ indicates a pair of neighboring sites, $(ij)=(ji)$ and 
\eqq{ 
J_{\rm s}^{\rm (HE)} = \sum_{l\neq 0} \frac{4 v_0^2}{l\Omega} |\mathcal{B}^{(l)}_{\bf e}|^2.  \label{eq:J_HE_1}
}
The spin operators are $\hat{{\bf s}} =\frac{1}{2}\sum_{\alpha,\beta=\uparrow,\downarrow} \hc^\dagger_\alpha \boldsymbol{\sigma}_{\alpha\beta} \hc_\beta$ with $\boldsymbol{\sigma}$ denoting the Pauli matrices.
HE stands for the high-frequency expansion. In equilibrium, $J_{\rm s}^{\rm (HE)}= \frac{4v_0^2}{U_0}>0$, since $\mathcal{B}^{(l_0)}=1$.

In a system with strong attractive interactions, the equilibrium state favors doublons or holons.
To express the low energy Hamiltonian, we introduce the pseudo-spins defined in the space of these local states as $\heta^+_i = (-)^i\hc^\dagger_{i\downarrow}  \hc^\dagger_{i\uparrow}$, $\heta^-_i = (-)^i\hc_{i\uparrow}  \hc_{i\downarrow}$ and $\heta^z_i =\frac{1}{2}(\hn_i-1)$. Here $(-)^i=1$ for the A sublattice and  $(-)^i=-1$ for the B sublattice. 
The low energy Hamiltonian can be expressed as 
\eqq{
\hH_{\rm dh} =   J_{\eta,XY}^{\rm (HE)}\sum_{(ij)}(\heta^x_i\heta^x_j +  \heta^y_i\heta^y_j) +  J_{\eta,Z}^{\rm (HE)} \sum_{(ij)} \heta^z_i\heta^z_j, \label{eq:H_eta}
}
where
\begin{subequations} \label{eq:J_HE_2}
\eqq{
J_{\eta,XY}^{\rm (HE)} &= - \sum_{l\neq 0} \frac{4 v_0^2}{l\Omega} \mathcal{B}^{(l)}_{\bf e}\mathcal{B}^{(l)}_{-\bf e}, \\
J_{\eta,Z}^{\rm (HE)} &= - \sum_{l\neq 0} \frac{4 v_0^2}{l\Omega} |\mathcal{B}^{(l)}_{\bf e}|^2.
}
\end{subequations}
In equilibrium, $J_{\eta,XY}^{\rm (HE)}=J_{\eta,Z}^{\rm (HE)}= -\frac{4v_0^2}{U_0}>0$, which can be connected to the spin SU$(2)$ symmetry in the repulsive Hubbard model via the Shiba transformation.
However, under the electric field, $J_{\eta,XY}^{\rm (HE)}\neq J_{\eta,Z}^{\rm (HE)}$ in general.\cite{Kitamura2016PRB}
We again note that these expressions are justified when $\Omega,|U| \gg v_0, |\Delta U|$.
For the following analysis, we introduce $I^{\rm (ex)}_1 = \sum_{l\neq 0} \frac{1}{l} |\mathcal{B}^{(l)}_{\bf e}|^2$ and $I^{\rm (ex)}_2 = \sum_{l \neq 0} \frac{1}{l} \mathcal{B}^{(l)}_{\bf e}\mathcal{B}^{(l)}_{-\bf e}$.

 %%%%%%%%%%%%%%%%%%%%%%%%%%%%%%%%%%%%%%%%%%%%%
 \begin{figure}[tb]
  \centering
    \hspace{-0.cm}
    \vspace{0.0cm}
\includegraphics[width=85mm]{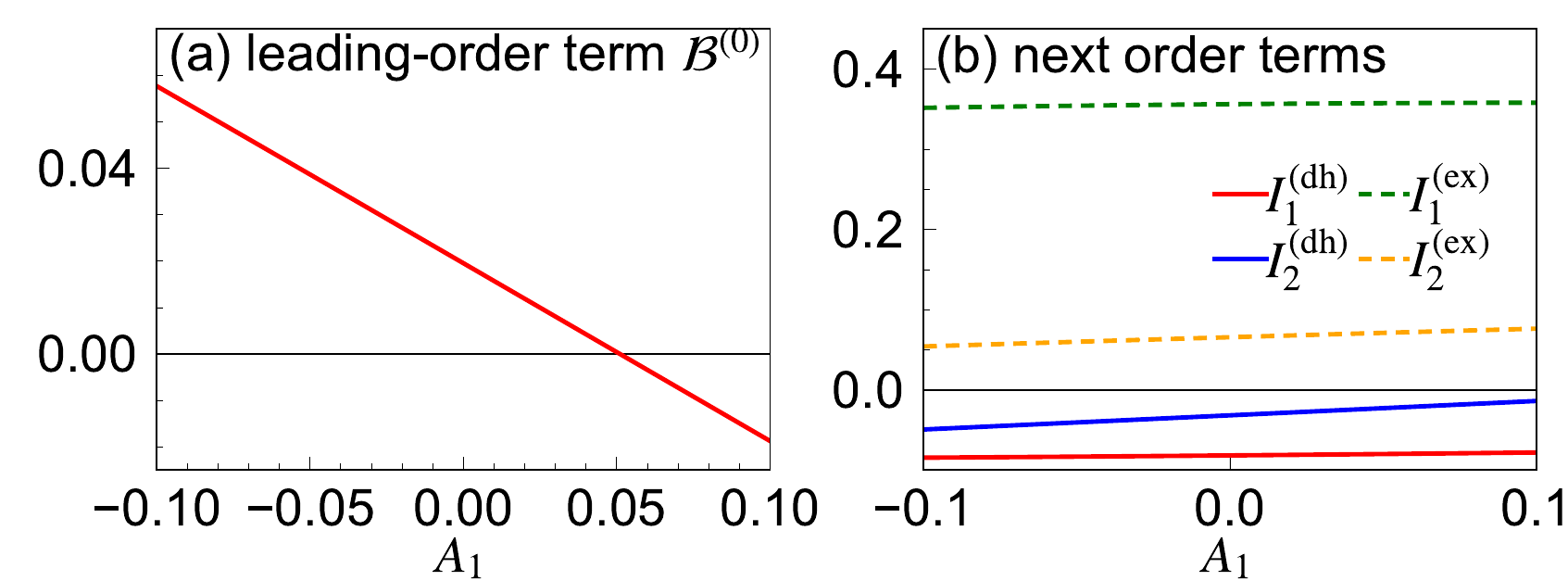} 
  \caption{(a) The coefficient of the d-h creation/annihilation term in the leading order Floquet Hamiltonian, i.e. $\mathcal{B}^{(0)}_{\bf e} = F^{(1)}[A_0,A_1,n_1,-l_0]$, as a function of $A_1$ for the type 1 protocol.
  (b) The coefficients of the next order terms in the Floquet Hamiltonian as a function of $A_1$ for the type 1 protocol.
  Here we use $A_{0}=-1$, $n_1 = 3$, $l_0=3$.}
  \label{fig:coeff_type1}
\end{figure}
%%%%%%%%%%%%%%%%%%%%%%%%%%%%%%%%%%%%%%%%%%%%

On the other hand, for $\Omega,|U|, |\Delta U|  \gg |v_0|$, we obtain slightly different expressions for the exchange couplings.
In this case, we consider the Floquet space (extended Hilbert space with photo-dressed states) and apply degenerate perturbation theory.~\cite{Mentink2015,Kitamura2016PRB}
For example, for $U>0$, starting from states with only singly occupied sites, we consider virtual excitations to states with a doublon-holon pair dressed with $l$ photons, 
where the transition amplitude is proportional to $\mathcal{A}^{(-l)}$.
As a result, we have $\hH_{\rm spin} = J_{\rm s}^{\rm (P)} \sum_{(ij)} \hat{\bf s}_i \cdot \hat{\bf s}_j$ with 
\eqq{
J_{\rm s}^{\rm (P)} = \sum_l \frac{4 v_0^2}{U-l\Omega} |\mathcal{A}^{(-l)}_{\bf e}|^2. \label{eq:J_ex_P}
}
Similarly, for $U<0$, we have $\hH_{\rm dh} =   J_{\eta,XY}^{\rm (P)}\sum_{(ij)}(\heta^x_i\heta^x_j +  \heta^y_i\heta^y_j) +  J_{\eta,Z}^{\rm (P)} \sum_{(ij)} \heta^z_i\heta^z_j$ with 
\begin{subequations} \label{eq:J_P_2}
\eqq{
J_{\eta,XY}^{\rm (P)} &= -\sum_l \frac{4 v_0^2}{U-l\Omega} \mathcal{A}^{(-l)}_{\bf e} \mathcal{A}^{(-l)}_{-\bf e}, \\
J_{\eta, Z}^{\rm (P)} &= -\sum_l \frac{4 v_0^2}{U-l\Omega} |\mathcal{A}^{(-l)}_{\bf e}|^2.
}
\end{subequations}
Note that these expressions break down at the resonance condition $U=l_0\Omega$, but without the diverging term $l=l_0$, $J^{\rm (P)}$ becomes equal to $J^{\rm (HE)}$.
The same expression can be obtained by introducing $\Omega_0$ such that $\Omega = k_0 \Omega_0$ and $U=l_0 \Omega_0$, where $k_0$ and $l_0$ have no common integer factor larger than 1.
More specifically, we consider the rotating frame of $e^{iUt}$ and apply the high-frequency expansion in terms of $\Omega_0$.\cite{Bukov2016PRL} 
For practical values of $|U|$ and off-resonant conditions, it is not a-priori clear whether Eqs.~\eqref{eq:J_HE_1},\eqref{eq:J_HE_2} or Eqs.~\eqref{eq:J_ex_P},\eqref{eq:J_P_2}  provide a better description of the low-energy properties.

%%%%%%%%%%%%%%%%%%%%%%%%%%%%%%%%%%
%%%%%%%%%%%%%%%%%%%%%%%%%%%%%%%%%%

\subsection{Assessment of the different protocols}
In this section, we evaluate $\mathcal{A}_{\bf e}^{(l)}(=\mathcal{A}_{-\bf e}^{(-l)*})$ for each excitation protocol and determine the condition for the suppression of the d-h creation terms in the Floquet Hamiltonian.

\subsubsection{Type 1 excitation} \label{sec:type1_ex}
In the type 1 protocol, in addition to the basic excitation by an electric field with frequency $\Omega$, we consider an electric field excitation of the form given in  Eq.~\eqref{eq:type1}. Then, 
we have 
\eqq{
\mathcal{A}_{\bf e}^{(l)}= \frac{1}{2\pi} \int^{2\pi}_0  e^{i [A_{0} \sin (\tau) + A_{1} \sin(n_1 \tau + \phi_1) -l\tau]} d\tau.
}
If $\phi_1=0$, $\mathcal{A}_{\bf e}^{(l)} $ is real, while without this condition, this is not guaranteed.
Therefore, in order to set the d-h creation term in the leading order to zero ($\mathcal{B}^{(0)}=0$), $\phi_1=0$ is favorable,
and we will focus on this case in the following.
Let us introduce the function
\eqq{
& F^{(1)}[A_0,A_1,n_1,l] \nonumber \\
& = \frac{1}{2\pi} \int^{2\pi}_0  \cos[A_{0} \sin (\tau) + A_{1} \sin(n_1 \tau) -l\tau] d\tau, \label{eq:demo1}
}
with $\mathcal{A}_{\bf e}^{(l)}= F^{(1)}[A_0,A_1,n_1,l]$.
Then $\mathcal{B}^{(0)}_{\bf e}=\mathcal{B}^{(0)}_{-\bf e}=0$ is equivalent to $F^{(1)}[A_{0},A_{1},n_1,-l_0]= F^{(1)}[-A_{0},-A_{1},n_1,-l_0]=0$.
To realize this, $n_1={\rm odd}$ is favorable, since 
\eqq{
& F^{(1)}[-A_0,-A_1,n_1,l]  = F^{(1)}[A_{0},A_{1},n_1,-l] \nonumber \\
& = F^{(1)}[A_0,(-)^{n_1+1}A_{1},n_1,l] (-)^l.
}
Intuitively, this condition can be associated with a parity argument. 
The auxiliary field $A_1$ provides additional excitation pathways at the energy of $n_1\Omega$. This is a parity-odd process.
On the other hand, the corresponding process using the $A_0$ field requires $n_1$ times the absorption of  the $\Omega$-frequency electric field excitation.
Thus, its parity is $(-)^{n_1}$. An inference can be expected between these processes if $n_1$ is odd.
Otherwise, the excitations involving these two processes end up with different final states.

 %%%%%%%%%%%%%%%%%%%%%%%%%%%%%%%%%%%%%%%%%%%%%
 \begin{figure}[tb]
  \centering
    \hspace{-0.cm}
    \vspace{0.0cm}
\includegraphics[width=85mm]{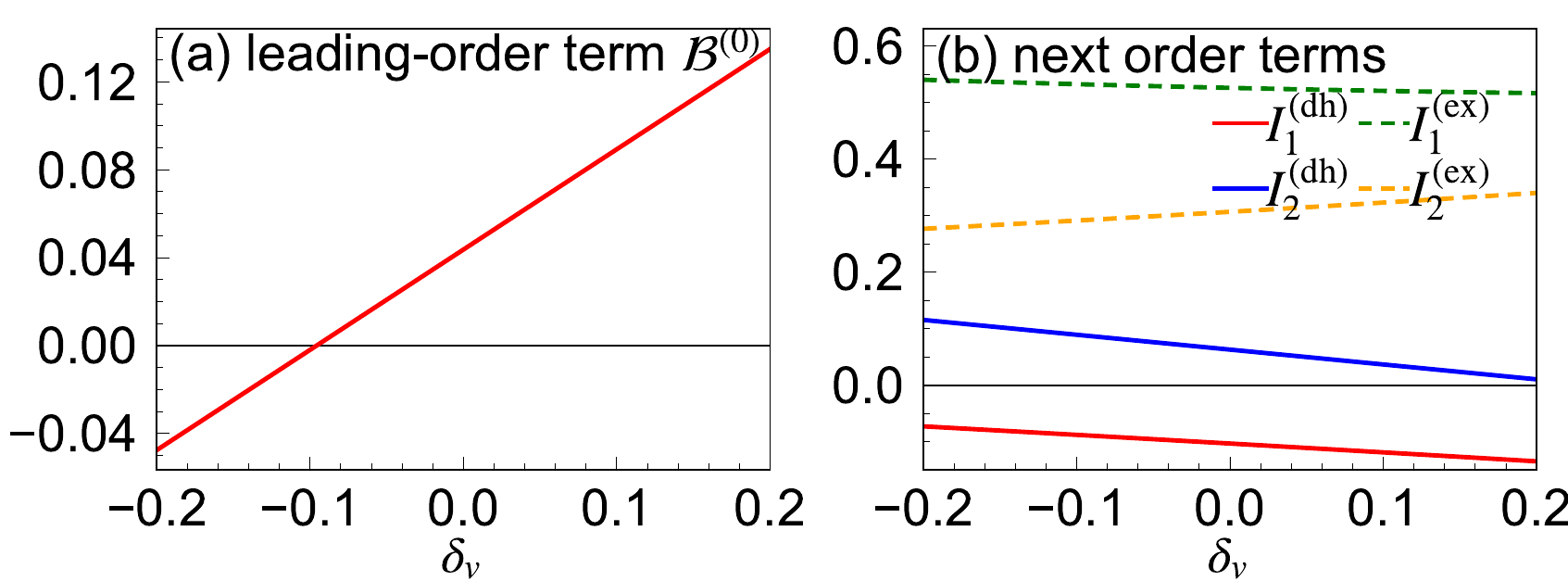} 
  \caption{(a) The coefficient of the d-h creation/annihilation term in the leading order Floquet Hamiltonian, i.e. $\mathcal{B}^{(0)}_{\bf e} = F^{(2)}[A_0,\delta_v,n_2,-l_0]$, as a function of $\delta_v$ for the type 2 protocol.
  (b) The coefficients of the next order terms in the Floquet Hamiltonian as a function of $\delta_v$ for the type 2 protocol.
  Here we use $A_0=-0.6$, $n_2 = 2$, $l_0=2$.}
  \label{fig:coeff_type2}
\end{figure}
%%%%%%%%%%%%%%%%%%%%%%%%%%%%%%%%%%%%%%%%%%%%

Since we have only one auxiliary field, the condition which sets $\mathcal{B}^{(0)}$ to zero determines the parameter of the auxiliary field.
We call this condition the ``optimal" condition for the type 1 protocol. Still, we note that it is not guaranteed that this protocol indeed produces the smallest heating, because of the contributions from other processes mentioned in the previous section.
In Fig.~\ref{fig:coeff_type1}, we illustrate how the coefficients of the effective model scale as a function of $A_1$.
Here, we take $A_0 = -1$, $n_1=3$ and $l_0=3$, which corresponds to the 3$\Omega$ absorption process which we mainly discuss in the following sections.
One can see that to minimize $|\mathcal{B}^{(0)}|$, $|A_1|$ much smaller than $A_0$ is sufficient.
In the perturbative regime with respect to the field strength, $\mathcal{B}^{(0)}$ scales as $\mathcal{O}(A_0^{l_0})$ for the type 0 excitation.
To cancel this term with $A_1$, we need $A_1=\mathcal{O}(A_0^{l_0})$.  
Thus, as long as $A_0$ remains close to the perturbative regime, we only need small $|A_1|$ to minimize $|\mathcal{B}^{(0)}|$.

\subsubsection{Type 2 excitation}\label{sec:type2_ex}

 %%%%%%%%%%%%%%%%%%%%%%%%%%%%%%%%%%%%%%%%%%%%%
 \begin{figure}[tb]
  \centering
    \hspace{-0.cm}
    \vspace{0.0cm}
\includegraphics[width=85mm]{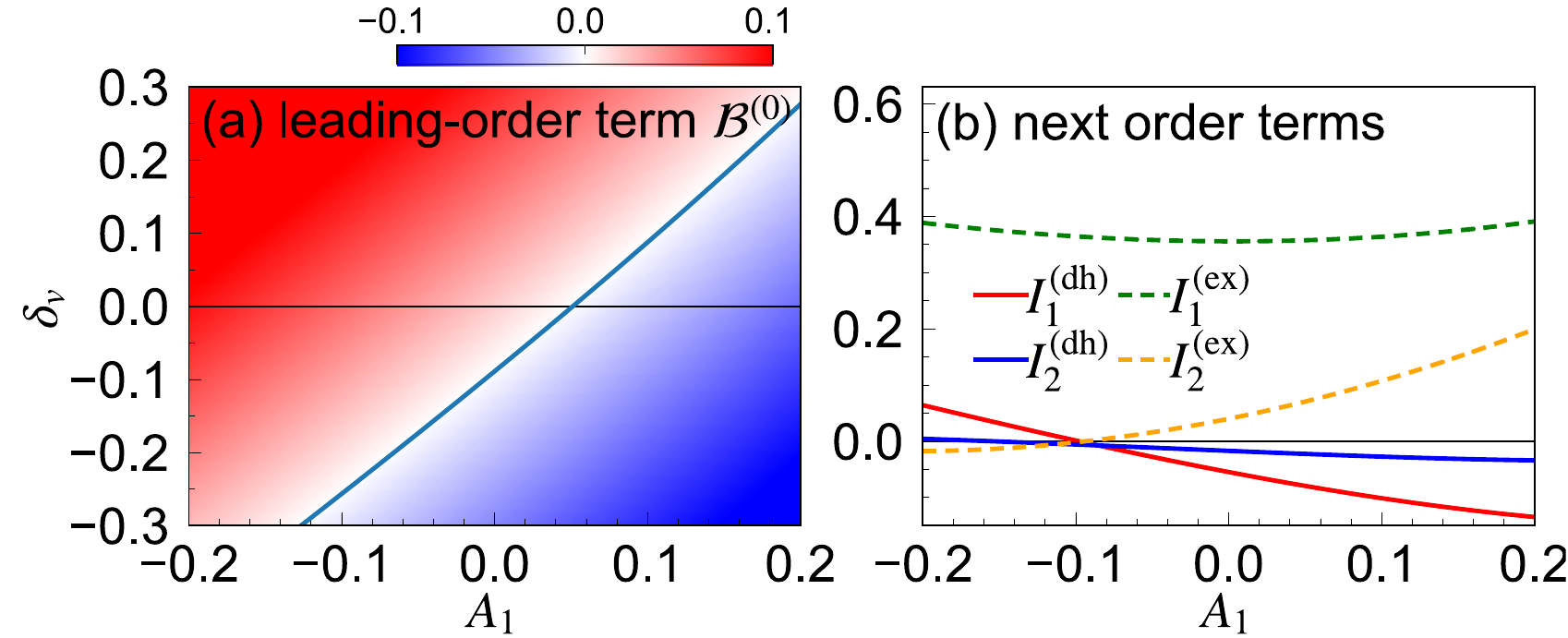} 
  \caption{
  (a) The coefficient of the d-h creation/annihilation term in the leading order Floquet Hamiltonian, $\mathcal{B}^{(0)}_{\bf e} = F^{(3)}[A_0,A_1,\delta_v, n_1,n_2,-l_0]$, as a function of $A_1$ and $\delta_v$ for the type 3 protocol.
  The solid blue line indicates the parameters that correspond to $\mathcal{B}^{(0)}_{\bf e} =0$.
  (b) The coefficients of the next order terms in the Floquet Hamiltonian as a function of $A_1$ for the type 3 protocol.
  We use $A_0=-1$, $n_1 = 3$, $n_2 = 2$, $l_0=3$.}
  \label{fig:coeff_type3}
\end{figure}
%%%%%%%%%%%%%%%%%%%%%%%%%%%%%%%%%%%%%%%%%%%%
In the type 2 protocol, on top of the basic electric field excitation, we consider an auxiliary hopping modulation of the form given in Eq.~\eqref{eq:type2}.
In this case, we have  $\mathcal{A}_{\bf e}^{(l)}=  \mathcal{J}_{l}(A_{0})  + \frac{\delta_v}{2} [e^{i\phi_2} \mathcal{J}_{l-n}(A_{0}) + e^{-i\phi_2} \mathcal{J}_{l+n}(A_{0})]$.
Here $\mathcal{J}_l$ is the $l$th order Bessel function.
If $\phi_2=0$, $\mathcal{A}_{\rm e}^{(l)} $ is real, while for other choices, this is not guaranteed.
Therefore, in order to make $\mathcal{B}^{(0)}_{\bf e}=\mathcal{B}^{(0)}_{-\bf e}=0$, $\phi_2=0$ is favorable, and we focus on this case in the following.
We introduce
\eqq{
F^{(2)}[A_{0},\delta_v,n_2,l] = \mathcal{J}_{l}(A_{0})  + \frac{\delta_v}{2} [\mathcal{J}_{l-n_2}(A_{0}) + \mathcal{J}_{l+n_2}(A_{0})] . \label{eq:demo2}
}
$\mathcal{B}^{(0)}=0$ corresponds to $F^{(2)}[A_{0},\delta_v,n_2,-l_0]= F^{(2)}[-A_{0},\delta_v,n_2,-l_0]=0$.
For this, $n_2={\rm even}$ is favorable, since 
\eqq{
& F^{(2)}[-A_{0},\delta_v,n_2,l]  = F^{(2)}[A_0,\delta_v,n_2,-l] \nonumber \\
 & = F^{(2)}[A_{0}, (-)^{n_2}\delta_v,n_2,l]  (-)^l.
}
This condition can be associated with a parity argument, as in the case of the type 1 excitation. The only difference is that the hopping process is an even parity process.

Since we have only one auxiliary field, the condition $\mathcal{B}^{(0)}=0$ determines the parameter of this auxiliary field.
We call this condition the ``optimal" condition for the type 2 protocol. 
In Fig.~\ref{fig:coeff_type2}, we illustrate how the coefficients of the effective model scale as a function of $A_1$.
Here, we take $A_0 = -0.6$, $n_2=2$ and $l_0=2$, which corresponds to the 2$\Omega$ absorption process discussed in the following sections.
Note that in the perturbative regime, $\delta_v$ for the optimal condition scales as $\mathcal{O}(A_0^{n_2})$.

\subsubsection{Type 3 excitation}

In the type 3 protocol, on top of the basic electric field excitation, we consider an auxiliary electric field excitation and an auxiliary hopping modulation as defined in Eq.~\eqref{eq:type3}. In other words,
this protocol is the combination of the type 1 and type 2 protocols. Here we set the phase shifts $\phi_1$ and $\phi_2$ to zero from the beginning. 
We introduce 
\eqq{
& F^{(3)}[A_0,A_1,\delta_v, n_1,n_2,l] \nonumber \\
& = \frac{1}{2\pi} \int^{2\pi}_0 [1+\delta_v \cos(n_2\tau)] e^{i(A_0\sin(\tau) + A_1\sin(n_1 \tau)} e^{-i \tau} d\tau \nonumber  \\
&= F^{(1)}[A_0,A_1,n_1,l]   \nonumber\\
&\;\;\;\;+ \frac{\delta_v}{2} \bigl\{F^{(1)}[A_{0},A_{1},n_1,l+n_2]  + F^{(1)}[A_{0},A_{1},n_1,l-n_2]  \bigl\} .
}
 Then we have $\mathcal{A}_{\bf e}^l = F^{(3)}[A_{0},A_{1},\delta_v, n_1,n_2,l]$. 
 This function satisfies 
 \eqq{
 & F^{(3)}[-A_0,-A_1,\delta_v, n_1,n_2,l]  = F^{(3)} [A_0,A_1,\delta_v, n_1,n_2,-l]\nonumber \\
 & = F^{(3)}[A_0,(-)^{n_1 + 1}A_1, (-)^{n_2}\delta_v, n_1,n_2,l] (-)^l.
 }
 Thus, in order to set the leading order terms to zero for all bonds, $n_1 = {\rm odd}$ and $n_2 = {\rm even}$ are favorable.
 
 In Fig.~\ref{fig:coeff_type3}(a), we show $\mathcal{B}^{(0)}$ in the plane of $A_1$ and $\delta_v$. 
 Here, we use $A_0 = -1, n_1=3, n_2=2$ and $l_0=3$, which corresponds to the 3$\Omega$ absorption process mainly discussed in the following sections.
 The solid blue line indicates the condition for $\mathcal{B}^{(0)}=0$.
  In Fig.~\ref{fig:coeff_type3}(b), we show the coefficients of the next order terms in the Floquet Hamiltonian as a function of $A_1$ along the $\mathcal{B}^{(0)}=0$ line in Fig.~\ref{fig:coeff_type3}(a).
  As can be seen, there is a point where both of $I^{(dh)}_1$ and $I^{(dh)}_2$ are suppressed, although they are not exactly zero.
  We choose the condition that minimizes $|I^{(dh)}_1|^2 + |I^{(dh)}_2|^2$ as the ``optimal" condition for the type 3 protocol.

%%%%%%%%%%%%%%%%%%%%%%%%%%%%%%%%%%
%%%%%%%%%%%%%%%%%%%%%%%%%%%%%%%%%%

\section{Results}\label{sec:results}

 %%%%%%%%%%%%%%%%%%%%%%%%%%%%%%%%%%%%%%%%%%%%%
 \begin{figure}[t]
  \centering
    \hspace{-0.cm}
    \vspace{0.0cm}
\includegraphics[width=65mm]{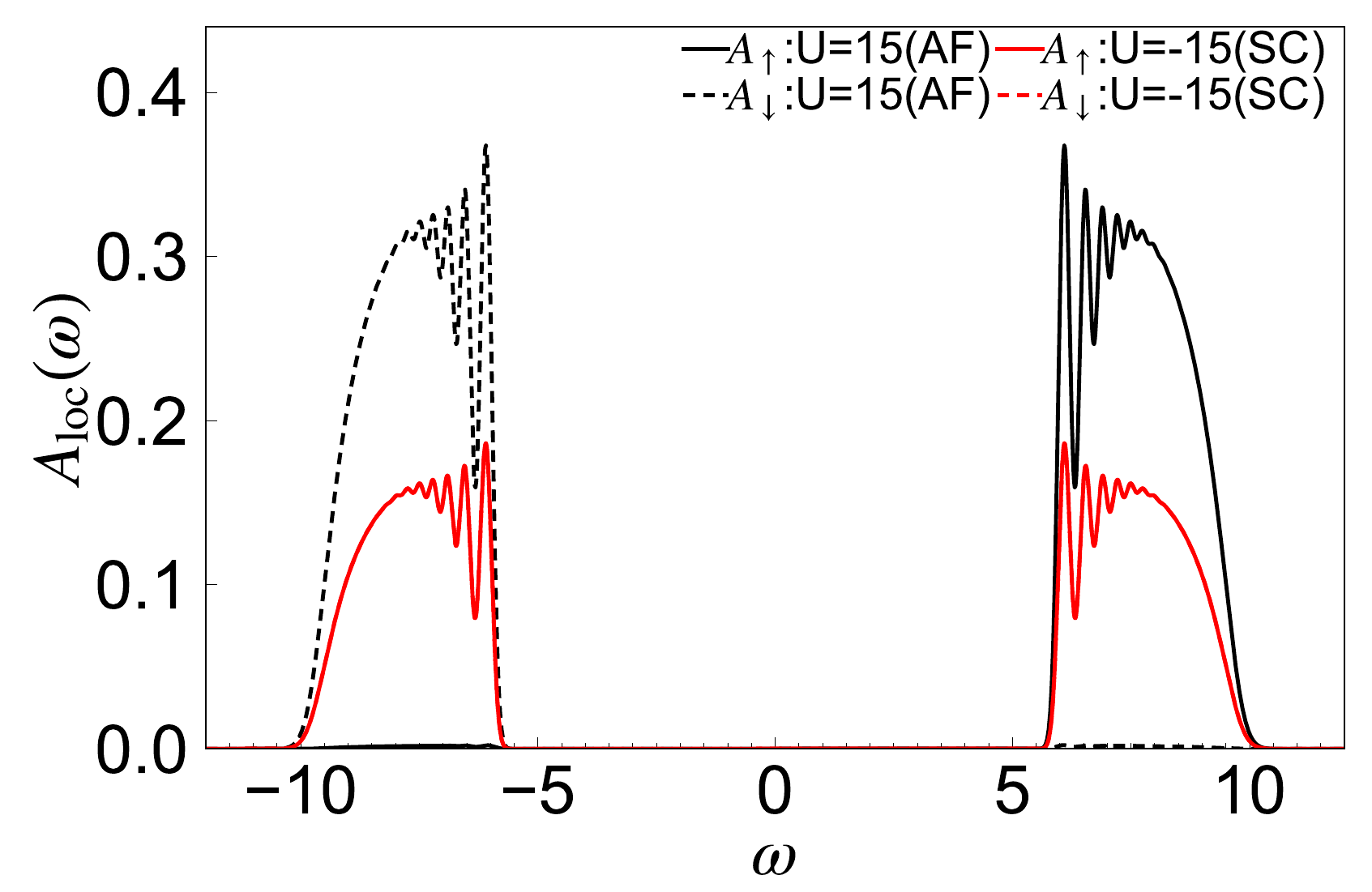} 
  \caption{Local single-particle spectral functions for $U=15$ and $U=-15$ in equilibrium at $T=0.02$. 
  For $U=15$, the system is in an AF state with spins aligned along the $z$ direction. 
  For $U=-15$, the system is in a SC state with a real order parameter. The red solid and dashed lines overlap.}
  \label{fig:Aw}
\end{figure}
%%%%%%%%%%%%%%%%%%%%%%%%%%%%%%%%%%%%%%%%%%%%

 %%%%%%%%%%%%%%%%%%%%%%%%%%%%%%%%%%%%%%%%%%%%%
 \begin{figure}[t]
  \centering
    \hspace{-0.cm}
    \vspace{0.0cm}
\includegraphics[width=87mm]{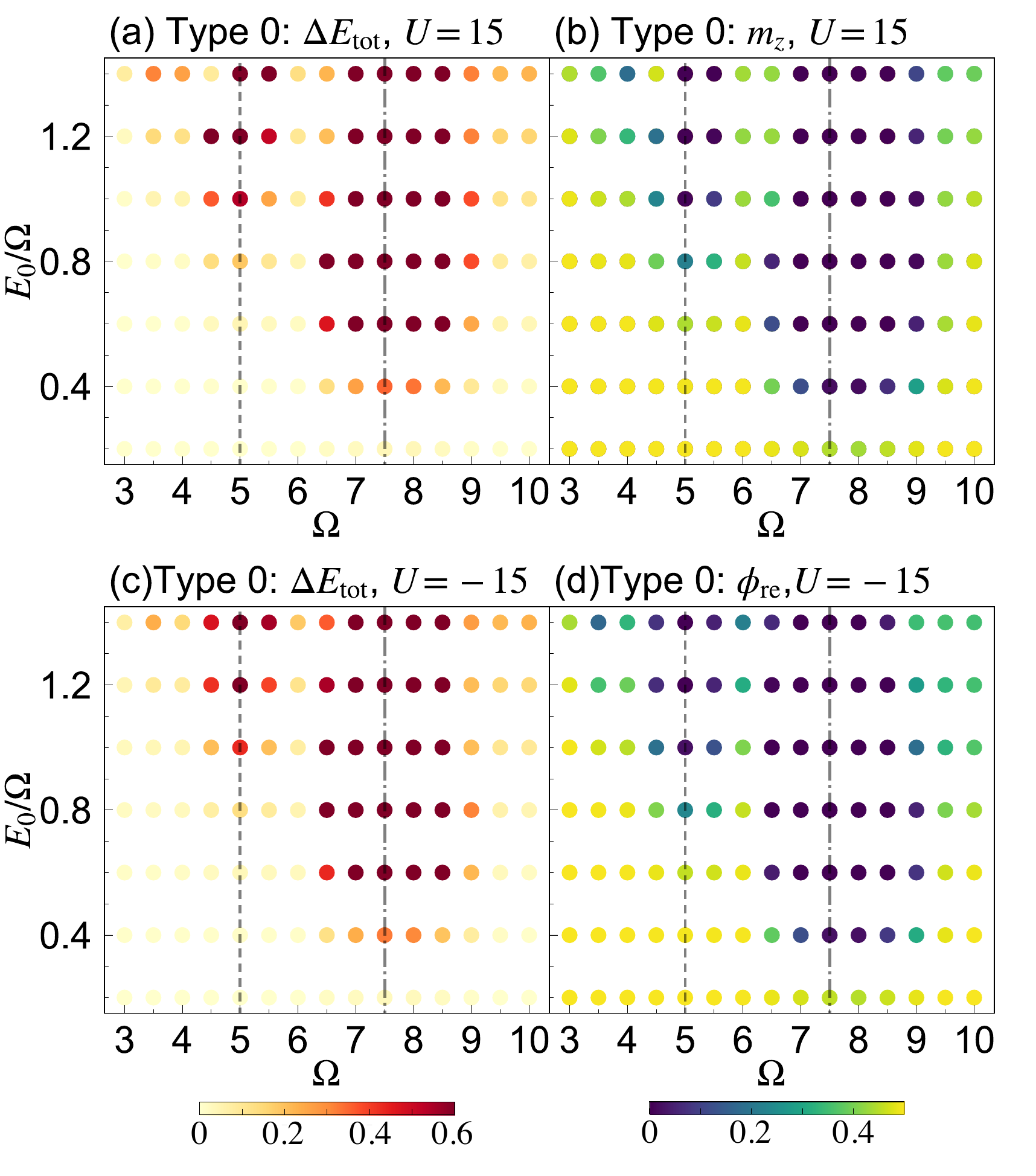} 
  \caption{The difference of the total energy from its equilibrium value ($\Delta E_{\rm tot}$) and the value of the order parameter ($m_z$ or $\phi_{\rm re}$) in the plane of the excitation frequency $\Omega$ and the field strength $E_0/\Omega$ $(=-A_0)$. These physical quantities are averaged over an interval of length $\frac{2\pi}{\Omega}$ around $t=50$, the type 0 excitation is used, and the initial temperature is $T=0.02$.
  (a) and (b) show results for $U=15$,  where the initial state is an AF state with spins aligned along the $z$ direction. 
  (c) and (d) show results for $U=-15$, where the initial state is a SC state with real order parameter. The vertical dashed lines indicate $\Omega = |U|/3$ and the vertical dot-dashed lines indicate $\Omega = |U|/2$.} 
  \label{fig:type0_global}
\end{figure}
%%%%%%%%%%%%%%%%%%%%%%%%%%%%%%%%%%%%%%%%%%%%

 %%%%%%%%%%%%%%%%%%%%%%%%%%%%%%%%%%%%%%%%%%%%%
 \begin{figure*}[t]
  \centering
    \hspace{-0.cm}
    \vspace{0.0cm}
\includegraphics[width=170mm]{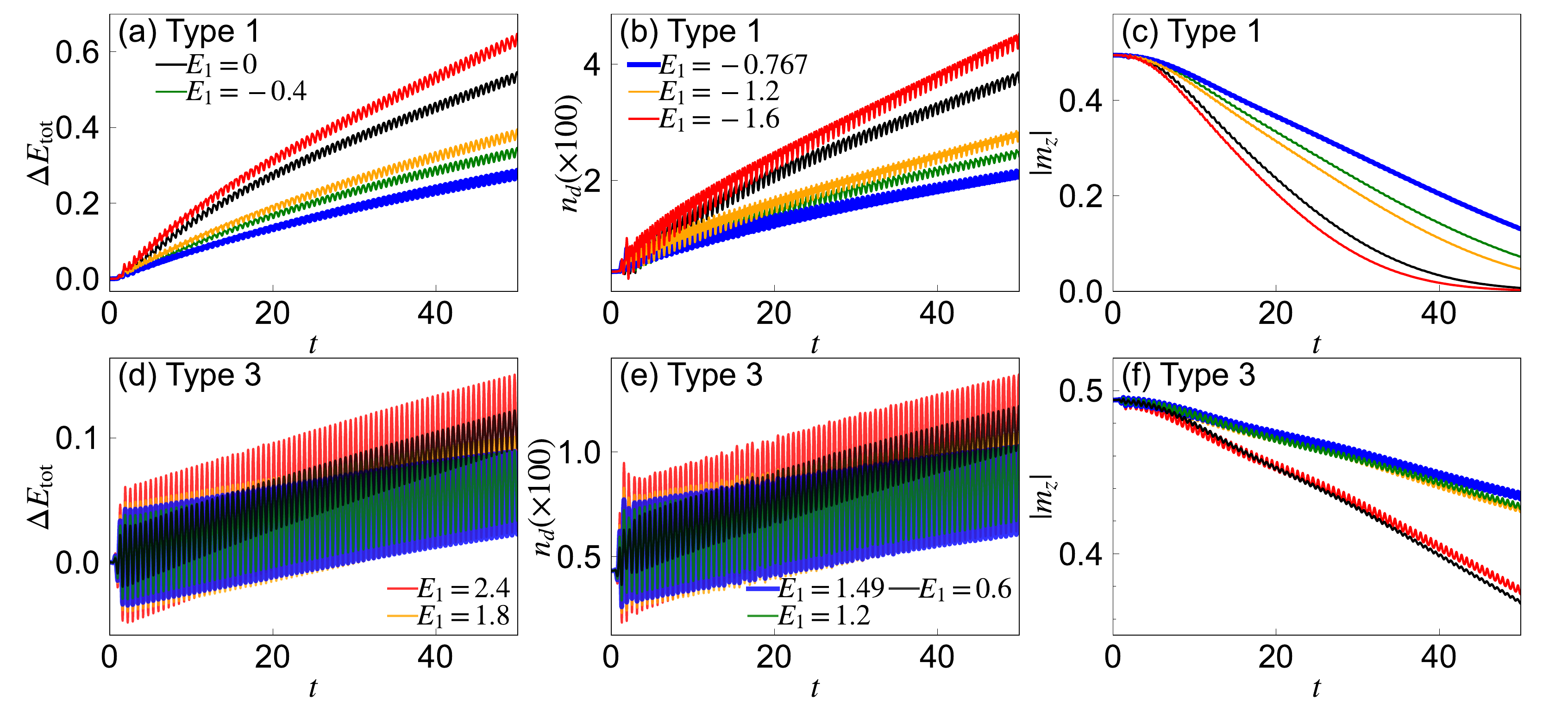} 
  \caption{(a)(d) Evolution of the difference of the total energy from its equilibrium value $\Delta E_{\rm tot}$, (b)(e) that of the number of doublons $n_d$, and  (c)(f) that of the staggered magnetization $m_z$ under periodic excitations with $\Omega=U/3$.
  Here, we set $U=15$ and the initial temperature to $T=0.02$, where the initial state is AF ordered with spins aligned along the $z$ direction. 
  The parameters of the main electric field excitation are $\Omega=5$ and $A_0=-1$ ($E_0=5$).
  For (a-c), we use the type 1 protocol with $l_0=3$ and $n_1=3$. These panels share the legend, which indicates the values of the auxiliary electric field.
 For (d-f),  we use the type 3 protocol with $l_0=3, n_1=3$ and $n_2=2$ and the strength of the auxiliary electric field and hopping modulation is chosen such that the d-h creation term vanishes in the leading order. These panels share the legend, and the parameters of the auxiliary fields are  $(E_1,\delta_v)=(0.6,-0.157),$ $(1.2,-0.223),$ $(1.49,-0.255),$ $(1.8,0.289),$ $(2.4,-0.354)$, but only the values of the auxiliary electric field are shown. The thick blue lines correspond to the optimal conditions for each protocol, as predicted by the Floquet Hamiltonian.  
  }
  \label{fig:U_3Ome_tevo}
\end{figure*}
%%%%%%%%%%%%%%%%%%%%%%%%%%%%%%%%%%%%%%%%%%%%

 %%%%%%%%%%%%%%%%%%%%%%%%%%%%%%%%%%%%%%%%%%%%%
 \begin{figure}[t]
  \centering
    \hspace{-0.cm}
    \vspace{0.0cm}
\includegraphics[width=70mm]{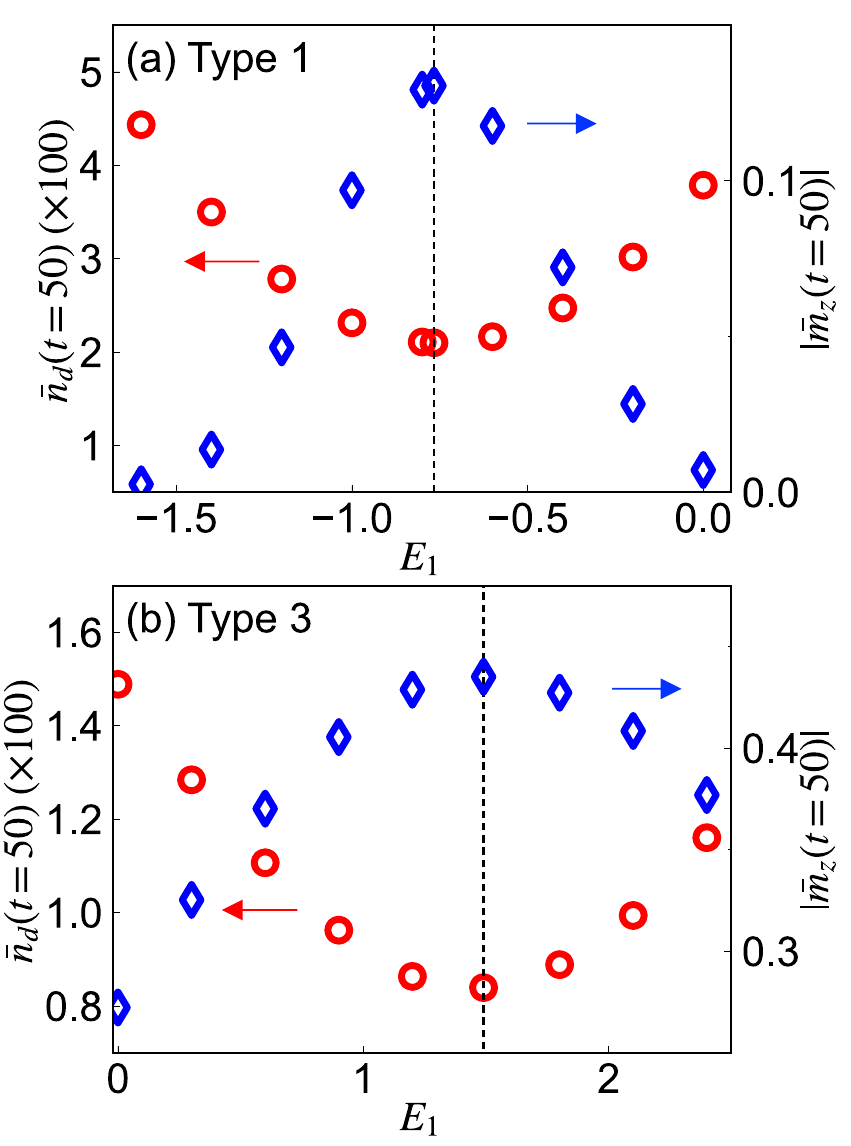} 
  \caption{ Number of doublons and staggered magnetization averaged over an interval $\frac{2\pi}{\Omega}$ around $t=50$ as a function of the strength of the auxiliary electric field ($E_1$).   We set $U=15$ and the initial temperature to $T=0.02$, where the initial state is AF. 
  The parameters of the main electric field excitation are $\Omega=5$ and $A_0=-1$. For (a), we use the type 1 protocol with $l_0=3$ and $n_1=3$.
 For (b),  we use the type 3 protocol with $l_0=3, n_1=3$ and $n_2=2$, and the strength of the auxiliary electric field and hopping modulation is chosen such that the d-h creation term vanishes in the leading order.
  The dashed lines indicate the optimal conditions for each protocol.
  }
  \label{fig:U_3Ome_tevo_summary}
\end{figure}
%%%%%%%%%%%%%%%%%%%%%%%%%%%%%%%%%%%%%%%%%%%%

\subsection{General remarks}

In this section, we analyze how well the suppression of heating with multi-color excitations works for the one-band Hubbard model.
To this end, we simulate the time evolution of the system under periodic excitations using nonequilibrium DMFT,~\cite{Georges1996,Aoki2013}
which becomes reliable in the limit of high spatial dimensions,
using the non-crossing approximation as the impurity solver.\cite{Eckstein2010b}
Our implementation is based on the open source library NESSi.\cite{Nessi2020} 
To reduce the computational cost of the simulations and enable systematic analyses, we consider the Bethe lattice in the following.
We set the bandwidth of the free particle ($U=0$) problem to $W=4$ and the interaction $|U|=15$.
%The interaction is chosen so that around the resonant condition of $|U|=3\Omega$ there is no another absorption with $l\Omega\neq 3\Omega$, but $\Omega$ is as small as possible.
%\textcolor{red}{The interaction is chosen such that $\Omega$ is smaller than the gap, but near the resonance condition $|U|=3\Omega$, there are no other relevant absorption processes with  $l\Omega\neq 3\Omega$.}
Note that we mainly focus on the sub-gap excitations, i.e. $\Omega$ is smaller than the gap. 
The interaction is chosen such that near the resonance condition $|U|=3\Omega$ ( $|U|=2\Omega$), there are no other relevant absorption processes with  $l\Omega\neq 3\Omega$ ($l\Omega\neq 2\Omega$),
but that $\Omega$ is reasonably small under this condition.
In the Bethe lattice, the effects of the electric field can be taken into account by considering two types of bonds connected to the effective impurity site, which are parallel and antiparallel to the external field, respectively. 
This treatment mimics the electric field applied to the body diagonal direction in hyper-cubic lattices.
Since the technical aspects have been discussed in previous works,\cite{Werner2017,Werner2018PRB,Jiajun2020PRB,Murakami2022PRL} we skip the details here.
In Appendix~\ref{sec:2d}, we show the results for the system on the two dimensional (2d) square lattice and
confirm that the Bethe lattice and the 2d square lattice produce the qualitatively same results.

\subsection{Equilibrium state at $t=0$}\label{sec:equilibrium}

One of the major goals of Floquet engineering is to control the low energy physics.
Thus, we choose as initial equilibrium states at $t=0$ ordered states and see how the order evolves under external fields at $t>0$.
In practice, we use $U=15$ or $U=-15$, and set the initial temperature low enough ($T=0.02$), so that the corresponding order parameters are almost saturated.
For $U=15$, the initial state is in the AF phase, with spins ordered along the $z$ direction.
The corresponding order parameter is $m_z(t) = \frac{1}{2}\langle \hn_{i\uparrow}(t) - \hn_{i\downarrow}(t) \rangle$.
For $U=-15$, we choose the initial state in the SC phase with real order parameter 
$\phi_{\rm re}(t) = {\rm Re} \langle \hc^\dagger_{i\uparrow}(t) \hc^\dagger_{i\downarrow}(t) \rangle$.
In Fig.~\ref{fig:Aw}, we show the local single-particle spectral functions of these states.
We plot $A_\sigma(\omega) = -\frac{1}{\pi} {\rm Im} G^R_{\sigma}(\omega)$, where $G^R_{\sigma}(\omega)$ is the Fourier transform
of $G^R_{\sigma}(t)\equiv -i\theta(t)\langle [\hc_{i,\sigma}(t),\hc^\dagger_{i,\sigma}(0)]_+\rangle$. 
Here, $\theta(t)$ is the Heaviside function, and $[\,,\,]_+$ indicates the anticommutator.
In both cases, the band gap is around $11$, while the separation between the bottom of the lower band and the top of the upper band is about $20$.
The peak structures in the upper and lower Hubbard bands in the AF phase correspond to spin-polarons which appear due to the spin-charge coupling in this system.\cite{Martinez1991PRB,Dagotto1994RMP,
Sangiovanni2006}
Since the AF phase for $U>0$ and the SC phase for $U<0$ are related by a particle-hole transformation (Shiba transformation), there are corresponding structures also in the spectrum of the SC phase.

\subsection{Suppression of heating}
Now we apply the external fields to the above mentioned initial states. 
In practice, we multiply the time periodic expressions for $\delta v(t)$ and $A(t)$, i.e. Eqs.~\eqref{eq:type0},\eqref{eq:type1},\eqref{eq:type2}, and \eqref{eq:type3},
by an envelope function $F(t;t_r)$ in order to smoothly switch the Hamiltonian from the equilibrium one to the time-periodic one within a time $t_r$.
The specific form of the ramp function is $F(t;t_r) = \theta(t_r-t )[\frac{1}{2}-\frac{3}{4}\cos(\pi t/t_r)+\frac{1}{4}\cos(\pi t/t_r)^3]  + \theta(t-t_r)$ and we set $t_r=2$.
Note that our choice of $t_r$ is short enough that it does not qualitatively affect the discussion below, which focuses on the dynamics under periodic excitations.

First, we show how the total energy and the order parameters evolve without auxiliary excitations.
In Fig.~\ref{fig:type0_global}, we show the difference of the total energy from its equilibrium value and the order parameters averaged around $t=50$ in the plane of the frequency $\Omega$ and the excitation strength $E_0/\Omega$ $(=-A_0)$.
Without excitations, the values of the order parameters are close to $0.5$.
In both the repulsive and attractive cases, the total energy resonantly increases around $|U|\simeq 3\Omega$ and $|U|\simeq 2\Omega$,
and, accordingly, the order parameters are reduced substantially.
The strong heating and melting of the order are caused by 3$\Omega$ and 2$\Omega$ absorption processes. 
Our goal is to suppress these absorption processes with auxiliary excitations.

 %%%%%%%%%%%%%%%%%%%%%%%%%%%%%%%%%%%%%%%%%%%%%
 \begin{figure}[t]
  \centering
    \hspace{-0.cm}
    \vspace{0.0cm}
\includegraphics[width=70mm]{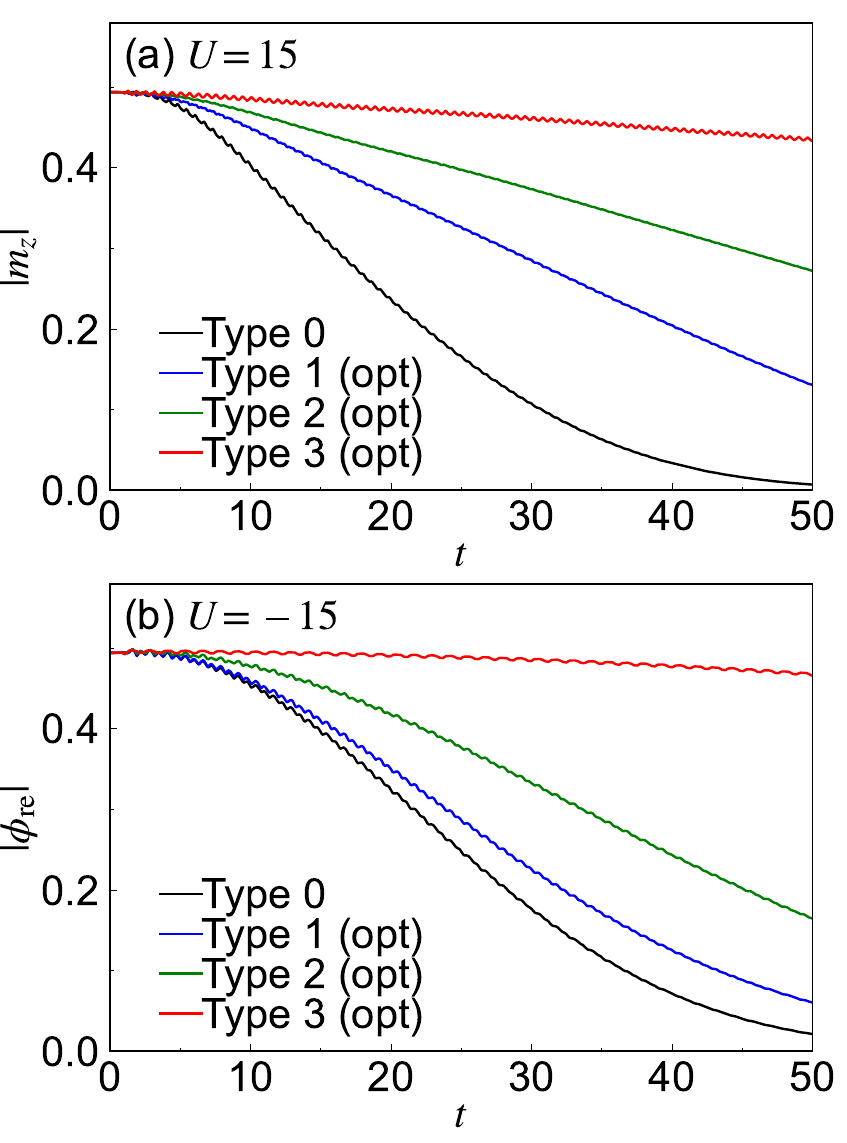} 
  \caption{Time evolution of the order parameters under the type 0, type 1, type 2 and type 3 protocols with optimal parameters. Here, $l_0=\pm 3$, $n_1=3$ and $n_2=2$. In panel (a), the system parameters and excitation conditions are the same as in Fig.~\ref{fig:U_3Ome_tevo}. 
  In panel (b), the excitation conditions are the same as in (a), but we use $U=-15$ and choose as the initial state the SC state with real order parameter at $T=0.02$.}
  \label{fig:3Ome_each_ef}
\end{figure}
%%%%%%%%%%%%%%%%%%%%%%%%%%%%%%%%%%%%%%%%%%%%

 %%%%%%%%%%%%%%%%%%%%%%%%%%%%%%%%%%%%%%%%%%%%%
 \begin{figure}[t]
  \centering
    \hspace{-0.cm}
    \vspace{0.0cm}
\includegraphics[width=85mm]{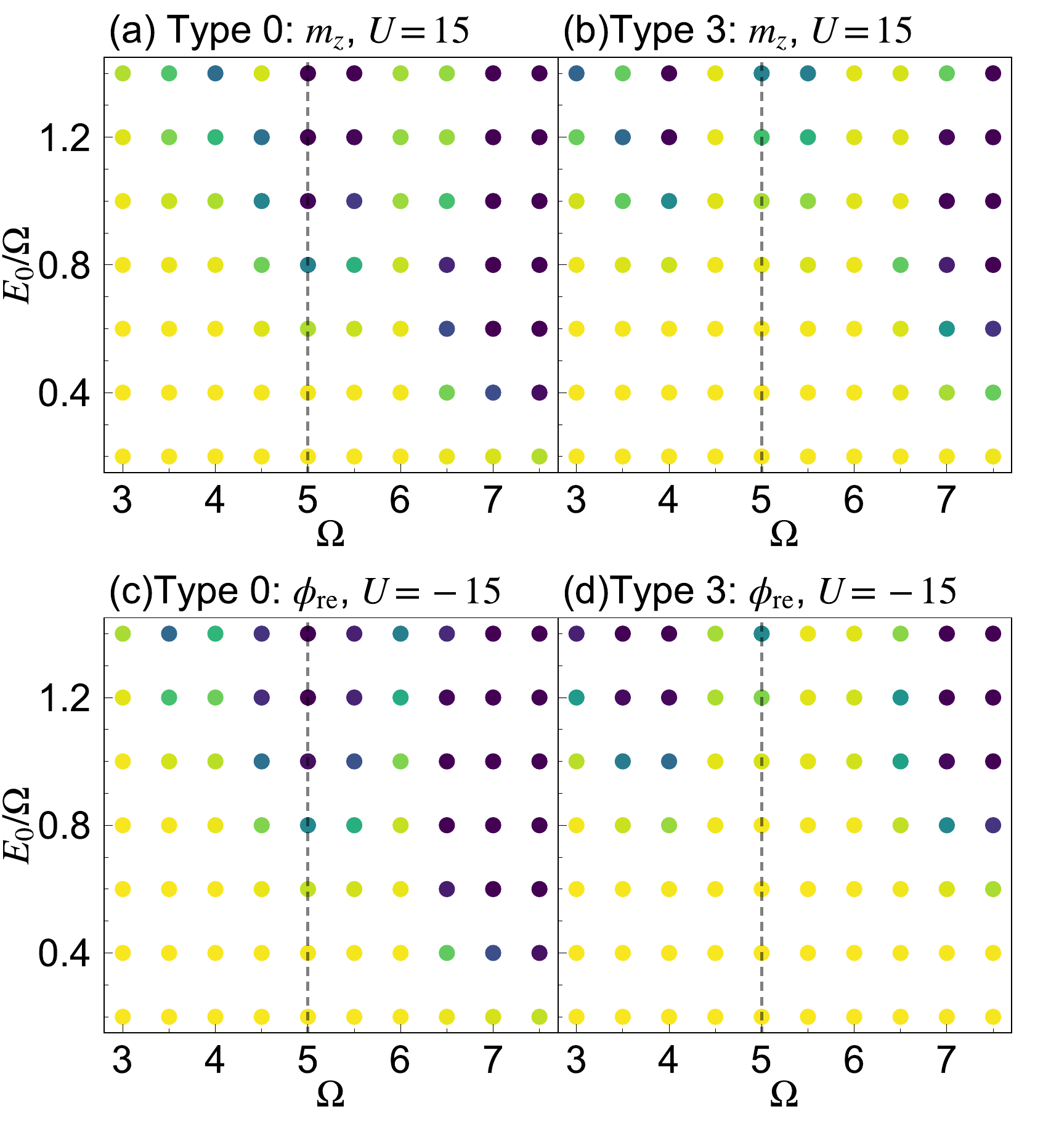} 
  \caption{The value of the order parameters ($m_z$ or $\phi_{\rm re}$) averaged around $t=50$ for (a)(b) $U=15$ and (c)(d) $U=-15$ in the plane of the excitation frequency $\Omega$ and the field strength $E_0/\Omega$ $(=-A_0)$.
  We focus on the regime around $|U|=3\Omega$. In panels (a) and (c), the type 0 excitation is used, while in (b) and (d), the type 3 excitation with the optimal parameters for $l_0=\pm 3$, $n_1=3$ and $n_2=2$ is used. The vertical dashed lines indicate $\Omega = |U|/3$.
  The other conditions and also the color code are the same as in Fig.~\ref{fig:type0_global}.}
  \label{fig:type1_global}
\end{figure}
%%%%%%%%%%%%%%%%%%%%%%%%%%%%%%%%%%%%%%%%%%%%

\subsubsection{Suppression of $3\Omega$ absorption}
We focus first on the 3$\Omega$ absorption and consider the excitation protocols with $l_0=\pm 3$, $n_1=3$ and $n_2=2$.
In Fig.~\ref{fig:U_3Ome_tevo}, we show the evolution of the difference of the total energy from its equilibrium value, $\Delta E_{\rm tot}$,  that of the number of doublons $n_d=\langle \hn_{i\uparrow} \hn_{i\downarrow}\rangle$, and that of the staggered magnetization $m_z$ for the type 1 and type 3 protocols. Here, we set $U=15, \Omega=5$ and $E_0=5$.
In Fig.~\ref{fig:U_3Ome_tevo_summary},  we plot $n_d$ and $m_z$ averaged around $t=50$ as a function of the strength of the auxiliary electric field $E_1$ for the type 1 and type 3 protocols.
In the case of the type 1 protocol, the increase of $\Delta E_{\rm tot}$ and $n_d$ as well as the melting of the AF order become slowest when the condition for $\mathcal{B}^{(0)}=0$ is met, i.e., for the ``optimal" condition for the type 1 protocol. 
Remember that we set the phase of the auxiliary field $\phi_1$ zero. We numerically confirmed that the heating and melting of order become stronger with nonzero $\phi_1$ (not shown), as we expected in Sec.~\ref{sec:type1_ex}.  
On the other hand, for the type 3 protocol, we change the parameters such that $\mathcal{B}^{(0)}=0$ is always satisfied.
The results show that the increase of  $\Delta E_{\rm tot}$ and $n_d$ and the melting of the AF order become slowest when the next-order d-h creation terms of the Floquet Hamiltonian are strongly suppressed, i.e., for the ``optimal" condition for the type 3 protocol.
In both protocols, $\Delta E_{\rm tot}$ and $n_d$ behave similarly, which indicates that in the present large-$U$ regime, the dominant heating process is the d-h creation.
The renormalization of the single-particle spectrum caused by the excitations is discussed in Appendix.~\ref{sec:Aw_floquet}.
The attractive model also shows the same behavior as Fig.~\ref{fig:U_3Ome_tevo} and Fig.~\ref{fig:U_3Ome_tevo_summary}.
Note that for the present parameters, the coefficients of the d-h creation terms in the leading and the next-leading order are comparable, see Fig.~\ref{fig:coeff_type1}. 
The success of the strategy of suppressing these terms order by order indicates that there is no significant interference between the processes described by the different terms in the effective Hamiltonian.
Also, the change in the coefficients of the d-h creation terms of higher orders should be moderate when we adjust the auxiliary field to suppress the low-order terms.

Let us note that a similar suppression of heating by multicolor driving is also observed for even larger $U$, see Appendix.~\ref{sec:U_scale}.
It can be well explained by the Floquet Hamiltonian how physical quantities for each protocol scale with $U$.
The fact that the Floquet Hamiltonian provides a good description indicates that the observed heating process can be associated with the Floquet prethermalization process.\cite{Herrmann_2017,Weidinger2017}
Higher order processes are expected to drive the system towards the infinite temperature state on longer timescales, where the description based on the Floquet Hamiltonian is no longer valid.\cite{Weidinger2017}

In Fig.~\ref{fig:3Ome_each_ef}(a)(b), we compare the evolution of the order parameters for the type 0 excitation, the optimal type 1 excitation and the optimal type 3 excitation.
Here, we also show the results for the optimal type 2 excitation for completeness. The optimal type 1 and type 2 excitations can be regarded as unoptimized cases for the type 3 excitation.
Although the effect of the auxiliary drive depends on the order, with the type 3 excitation, melting of the order is strongly suppressed both in the repulsive and attractive models.
We also note that, for the type 2 excitation, one finds that the optimal condition predicted by the Floquet Hamiltonian ($\delta_v=-0.0889$ in the case of Fig.~\ref{fig:3Ome_each_ef}) indeed leads to an efficient suppression of heating
both for the AF and SC phases. However, for the SC phase, it turns out that heating can be further suppressed by decreasing $\delta_v$ from $-0.0889$ (not shown).
This may be because both $I^{\rm (dh)}_1$ and $I^{\rm (dh)}_2$ are further reduced simultaneously, unlike in the case of the type 1 protocol, see Fig.~\ref{fig:coeff_type1}.

 %%%%%%%%%%%%%%%%%%%%%%%%%%%%%%%%%%%%%%%%%%%%%
 \begin{figure}[t]
  \centering
    \hspace{-0.cm}
    \vspace{0.0cm}
\includegraphics[width=87mm]{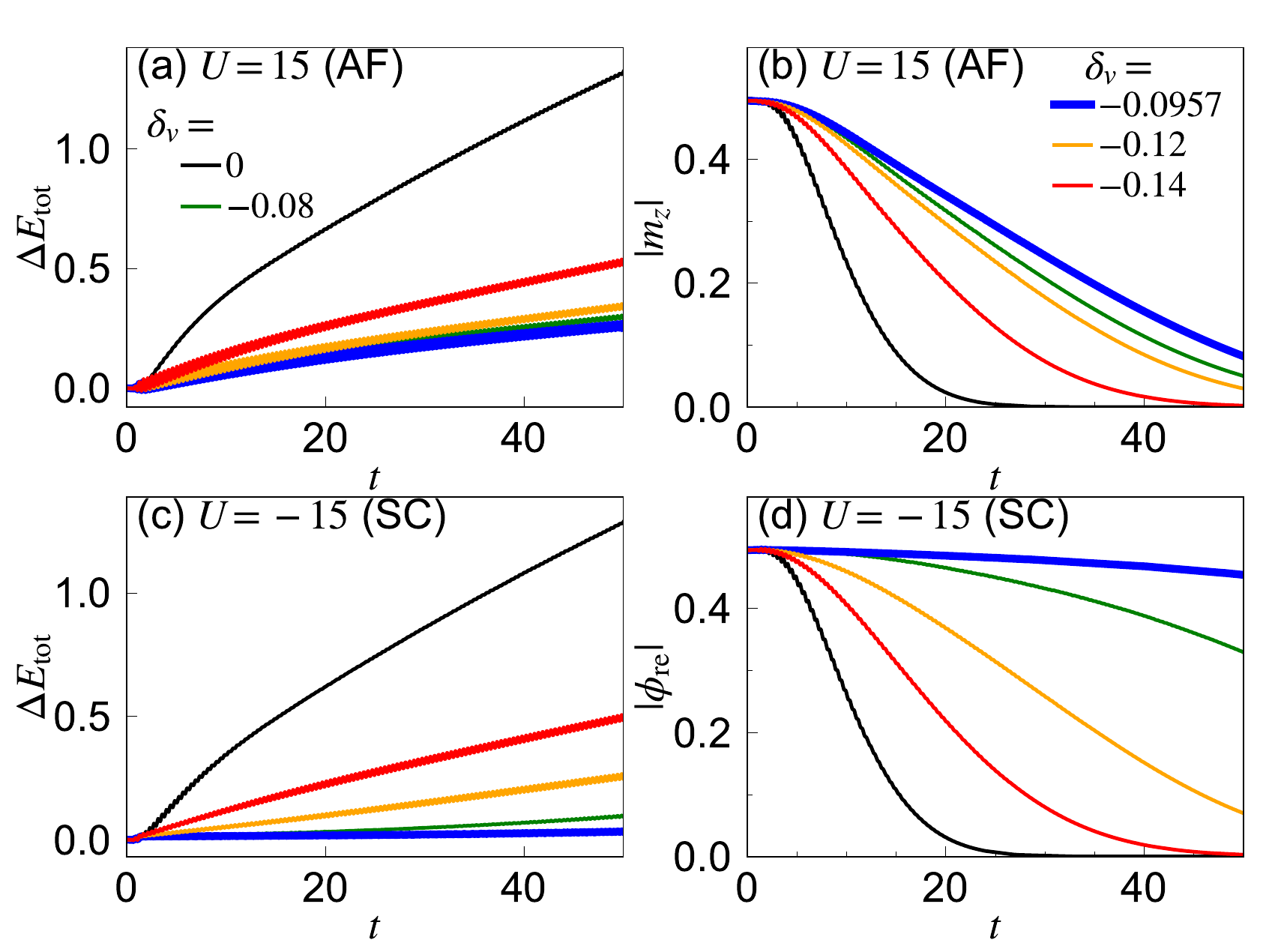} 
  \caption{(a)(c) Time evolution of the deviation of the total energy from its equilibrium value $\Delta E_{\rm tot}$ and (b)(d) that of the order parameters for type 2 protocols with $l_0 = \pm 2$, $n_2=2$ and the indicated values of $\delta_v$.
  The parameters of the main drive are $\Omega=7.5$ and $A_0=-0.6$ ($E_0=4.5$).
  All panels share the legend.
 In (a) and (b), we set $U=15$ and the initial temperature to $T=0.02$, so that the initial state is AF with spins aligned along the $z$ direction. 
  In (c) and (d), we set $U=-15$ and the initial temperature to $T=0.02$, where the initial state is a SC state with real order parameter. 
  The thick blue lines correspond to the optimal conditions predicted by the Floquet Hamiltonian.
  }
  \label{fig:U_2Ome_tevo}
\end{figure}
%%%%%%%%%%%%%%%%%%%%%%%%%%%%%%%%%%%%%%%%%%%%

 %%%%%%%%%%%%%%%%%%%%%%%%%%%%%%%%%%%%%%%%%%%%%
 \begin{figure}[t]
  \centering
    \hspace{-0.cm}
    \vspace{0.0cm}
\includegraphics[width=85mm]{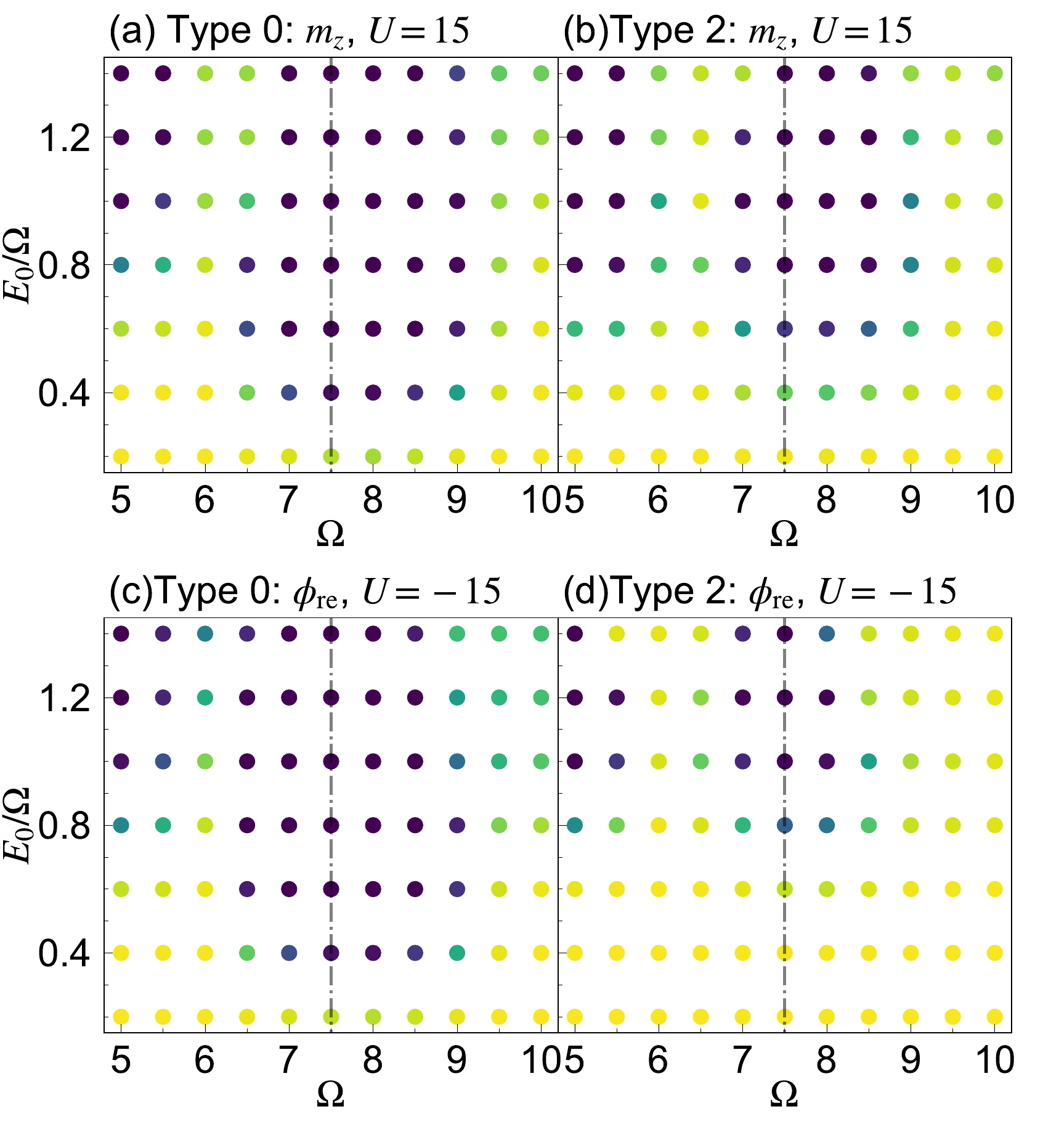} 
  \caption{The values of the order parameters ($m_z$ or $\phi_{\rm re}$) averaged around $t=50$ for (a)(b) $U=15$ and (c)(d) $U=-15$ in the plane of the excitation frequency $\Omega$ and the field strength $E_0/\Omega$ $(=-A_0)$.
  We focus on the regime around $|U|=2\Omega$. In panels (a) and (c), the type 0 excitation is used, while in (b) and (d), the type 2 excitation with the optimal parameters for $l_0=\pm 2$ and $n_2=2$ is used.
  The vertical dot-dashed lines indicate $\Omega = |U|/2$.
  The other conditions and the color code are the same as in Fig.~\ref{fig:type0_global}.}
  \label{fig:type2_global}
\end{figure}
%%%%%%%%%%%%%%%%%%%%%%%%%%%%%%%%%%%%%%%%%%%%

In Fig.~\ref{fig:type1_global}, we compare the values of order parameters averaged around $t=50$ obtained by the type 0 and type 3 protocols.
For the type 3 calculations, the optimal parameters are used.
Around $\Omega=5$, the melting of the order is strongly suppressed by the auxiliary drives for both $U=15$ and $U=-15$,
which can be attributed to the suppression of the $3\Omega$ absorption as discussed above.
Around $\Omega=6\sim8$, one can also observe a slower melting. 
However in this regime, $2\Omega$ absorption can also happen.
To fully discuss the conditions for the suppression of the $2\Omega$ absorption, one needs to consider the effective model in the $\pm 2\Omega$ rotating frame.
Around $\Omega=4$, the melting speed is increased, which may be attributed to the enhancement of the $4\Omega$ absorption processes by the auxiliary excitations.
We note that the behavior of the total energy is almost the same as that of the order parameter (not shown). More specifically, the suppression of the order parameters is accompanied by an increase of the total energy.
These results demonstrate that the Floquet Hamiltonian for a given rotating frame serves as a useful guide for protocols which suppress the corresponding absorption via multi-color excitations, but auxiliary excitations can both suppress or enhance other absorptions.

\subsubsection{Suppression of $2\Omega$ absorption}
Next, we try to suppress the 2$\Omega$ absorption using the type 2 protocol with $l_0=\pm 2$ and $n_2=2$.
In Fig.~\ref{fig:U_2Ome_tevo}, we show the evolution of the order parameters and of the deviation of the total energy from its equilibrium value, both for $U>0$ and $U<0$ and for the indicated values of $\delta_v$.
Here, we use $|U|=15$, $\Omega=7.5$, and $A_0 = -0.6$.
The rate of increase in the total energy and the speed of melting of the order parameters are slowest when the leading order d-h creation terms become zero, i.e., when $\mathcal{B}^{(0)}=0$.
The suppression of heating is more effective for the SC phase than for the AF phase.
We note that the next leading dh creation/annihilation terms remain nonzero in the optimal condition, see Fig.~\ref{fig:coeff_type2}.
The difference between AF and SC should be attributed to how these terms act on the different ordered phases.
Remember that we set the phase of the auxiliary field $\phi_2$ to zero. We numerically confirmed that the heating and melting of the orders becomes stronger with nonzero $\phi_2$ (not shown), as we expected in Sec.~\ref{sec:type2_ex}. 
We summarize the results in Fig.~\ref{fig:type2_global}, where we compare the values of order parameters at $t=50$ around $|U|\simeq 2\Omega$ between the type 0 protocol and the type 2 protocol with the ``optimal" parameters.
For both $U=15$ and $U=-15$, the 2$\Omega$ absorption is suppressed. As expected from Fig.~\ref{fig:U_2Ome_tevo}, the suppression is strong for $U<0$ (the SC phase).
In addition, we checked that the behavior of the total energy is almost the same as that of the order parameter (not shown).

%%%%%%%%%%%%%%%%%%%%%%%%%%%%%%%%%%
%%%%%%%%%%%%%%%%%%%%%%%%%%%%%%%%%%
\subsection{Extraction of exchange couplings}

 %%%%%%%%%%%%%%%%%%%%%%%%%%%%%%%%%%%%%%%%%%%%%
 \begin{figure}[t]
  \centering
    \hspace{-0.cm}
    \vspace{0.0cm}
\includegraphics[width=80mm]{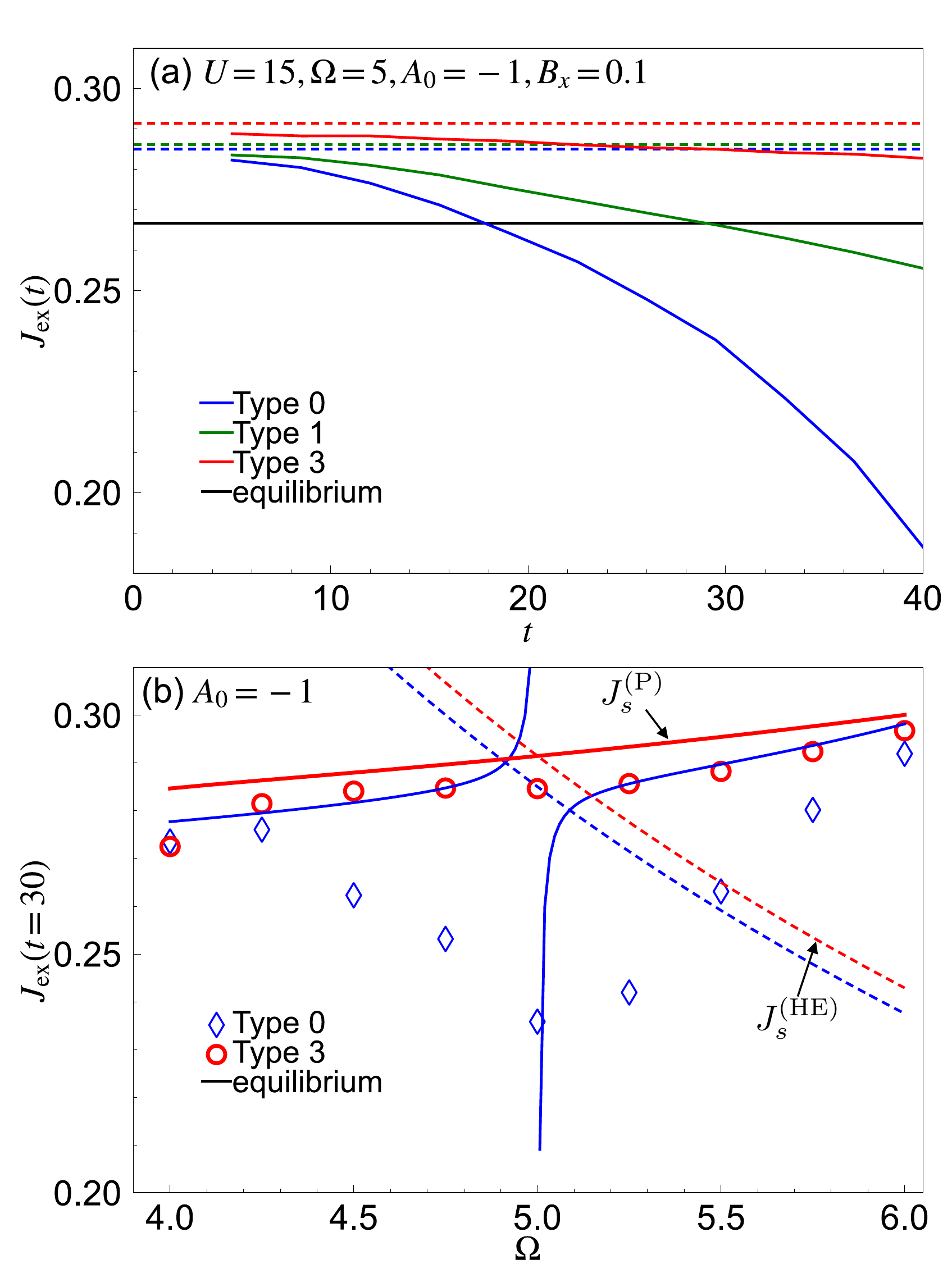} 
  \caption{(a) Time evolution of the exchange coupling evaluated from Eq.~\eqref{eq:J_ex_rep} for $U=15,\Omega=5,A_0=-1$ and $B_x=0.1$. The dashed lines indicate $J^{\rm (HE)}_s$  (Eq.~\eqref{eq:J_HE_1}),
  while the solid black line indicates the equilibrium value of the exchange coupling. 
  Here, we set $l_0=n_1 =3$ and $n_2=2$.
  (b) The estimated exchange couplings as a function of $\Omega$ at $t=30$ for $U=15$, $A_0=-1$ and $B_x=0.1$. 
  Here the type 3 excitation with the optimal parameters for $l_0=n_1 =3$ and $n_2=2$ is used.  
  The dashed lines show $J^{\rm (HE)}_s$ (Eq.~\eqref{eq:J_HE_1}), while colored solid curves show $J^{\rm (P)}_s$ (Eq.~\eqref{eq:J_ex_P}). The horizontal black line indicates the equilibrium value. }
  \label{fig:J_ex_repulsive}
\end{figure}
%%%%%%%%%%%%%%%%%%%%%%%%%%%%%%%%%%%%%%%%%%%%

In this section we illustrate how the suppression of heating helps us to observe the modification of the low energy physics due to the virtual excitations induced by the periodic excitations.
In the present systems, this low energy physics is characterized by the exchange couplings between the spins or the pseudo-spins  in 
the repulsive and the attractive regimes, respectively.
As discussed in previous works,\cite{Eckstein2014PRL,Mentink2015} one way to measure the exchange couplings is to compare the (pseudo-)spin dynamics obtained from DMFT  with the mean-field (MF) dynamics of the Heisenberg (XXZ) spin models, i.e. the solution of a Landau-Lifshitz (LL) equation.

We first discuss how the measurement of the exchange coupling works for repulsive $U$. 
Note that the periodic excitation does not break the SU(2) symmetry.
When the exchange coupling is modified by the periodic excitations, a spin precession occurs if we additionally apply a homogeneous magnetic field.
If we assume that the homogeneous magnetic field is applied along the $x$ direction, the spin model can be written as $\hH_{\rm spin}(t) = J_{\rm ex}(t) \sum_{( i,j)} \hat{\bf s}_i\hat{\bf s}_j + B_x {\bf e}_x \sum_i \hat{\bf s}_i$.
The MF Hamiltonian becomes $\hH^{\rm MF}_{\rm spin} = \sum_i {\bf B}_i^{\rm eff} {\bf \hat{s}}_i $ where $ {\bf B}_i^{\rm eff}=J_{\rm ex}(t)\sum_\delta {\bf s}_{i+\delta}(t) + B_x {\bf e}_x$.
Here, ${\bf s}(t)$ is the expectation value of $\hat{\bf s}$ and $\delta$ indicates the neighboring sites. Then the corresponding LL equation becomes $\partial_t {\bf s}_i(t) = {\bf B}_i^{\rm eff}(t) \times  {\bf s}_i(t)$.
In the present case, we have the relation $s_{\text{A},y,z}= -s_{\text{B},y,z}$ and $s_{\text{A},x}= s_{\text{B},x}$ (A and B are sublattice indices). From this we obtain 
\eqq{
ZJ_{\rm ex} &= -\frac{B_0}{2s_x} - \frac{\dot{s}_y}{2s_xs_z}. \label{eq:J_ex_rep}
}
Here, $\dot{s}_y$ is the time derivative, $Z$ is the number of neighboring sites and we omit the time indices. In the previous works, this equation has been used to discuss the effects of photo-doping~\cite{Eckstein2014PRL} or periodic excitations on the exchange interactions.\cite{Mentink2015} In this study, we apply the above formula to the time-averaged values of ${\bf s}(t)$, $\bar{\bf s}(t) \equiv \frac{1}{T_p}\int^{t+T_p/2}_{t-T_p/2} {\bf s}(t) dt$, where $T_p=\frac{2\pi}{\Omega}$, to eliminate effects related to the micro-motion.

In Fig.~\ref{fig:J_ex_repulsive}(a), we show the resultant values of $J_{\rm ex}(t)$ for different excitation protocols for $U=15$, $\Omega=5$ and $A_0=-1$. 
For this parameter set, the Floquet Hamiltonian produces an enhanced $J_{\rm ex}(t)$, compared to the equilibrium value, due to the virtual excitations associated with the periodic excitations.
For the type 0 protocol, the value initially matches $J_{s}^{\rm (HE)}$, but $J_{\rm ex}(t)$ gradually decreases with time.
For the type 1 protocol, the tendency is the same, but the deviation from the expected exchange coupling is less severe. 
For the type 3 protocol, $J_{\rm ex}(t)$ remains close to the expected coupling within the time range studied here.
The deviation from the expected value can be attributed to the photo-doping. 
In Ref.~\onlinecite{Eckstein2014PRL}, it was pointed out that photo-doping effectively reduces $J_{\rm ex}$, similar to chemical doping. 
This behavior can be intuitively understood since the doping reduces the number of singly occupied sites, so that the probability of finding spins on neighboring sites which develop correlations is reduced.\cite{Murakami2023}
With the type 0 protocol, as time evolves, more particles are excited across the gap to reduce $J_{\rm ex}$, which competes with the effects of the virtual excitations.
The undesired photo-doping (absorption) can be suppressed with the type1 and type 3 protocols.

To summarize, in Fig.~\ref{fig:J_ex_repulsive}(b), we show how the estimated $J_{\rm ex}$ behaves at $t=30$ near the $U= 3\Omega$ resonance.
Let us first discuss the behavior of the type 0 protocol.
For this protocol, the prediction from the perturbation theory $J_{\rm ex}^{(\text{P})}$ shows a divergence at $U= 3\Omega$ due to the virtual fluctuations to the $3\Omega$-dressed states,
while the prediction from the high-frequency expansion, $J_{\rm ex}^{(\text{HE})}$, is regular since the contribution from such states is not included.
However, both fail to explain the actual behavior of $J_{\rm ex}$,
which is strongly suppressed near the resonance $U=3\Omega$.
This can be attributed to the strong enhancement of the density of photo-carriers in this regime, which is neither taken into account in  $J_{\rm ex}^{(\text{HE})}$ nor in $J_{\rm ex}^{\rm(P)}$.
As for the type 3 protocol,  $J_{\rm ex}$ is generally larger than the corresponding value for the type 0 protocol. 
The resonant reduction of $J_{\rm ex}$ around the resonance is absent, and $J_{\rm ex}$ tends to increase with increasing $\Omega$.
For the type 3 protocol, $J_{\rm ex}^{\rm(P)}$ is regular as a function of $\Omega$, since the diverging term is suppressed due to the condition $\mathcal{B}^{(0)}=0$.
$J_{\rm ex}^{(\text{HE})}$ is also regular. Note that $J_{\rm ex}^{(\text{HE})}$ becomes equal to $J_{\rm ex}^{\rm(P)}$ for the type 3 protocol at $U=3\Omega$.
Importantly, the behavior of the numerically evaluated $J_{\rm ex}$ is well explained by $J_{\rm ex}^{\rm(P)}$ for the type 3 protocol.

Remember that the conditions under which $J_{\rm ex}^{\rm(P)}$ and $J_{\rm ex}^{(\text{HE})}$ are justified are different.  
$|U|,\Omega,|\Delta U|\gg |v_0|$ is required for $J_{\rm ex}^{\rm(P)}$, while $|U|,\Omega\gg |v_0|, |\Delta U|$ is required for $J_{\rm ex}^{(\text{HE})}$.
The results show that under the condition that heating is well suppressed, $J_{\rm ex}^{\rm(P)}$ describes the low energy physics better than $J_{\rm ex}^{(\text{\text{HE}})}$ in practice.
The qualitative difference between the $\Omega$ dependence of $J_{\rm ex}^{(\text{HE})}$ and $J_{\rm ex}^{\rm(P)}$
can be attributed to additional contributions from the $\Delta U$ term in the denominator in  $J_{\rm ex}^{\rm(P)}$, which are absent in $J_{\rm ex}^{(\text{HE})}$.

 %%%%%%%%%%%%%%%%%%%%%%%%%%%%%%%%%%%%%%%%%%%%%
 \begin{figure}[t]
  \centering
    \hspace{-0.cm}
    \vspace{0.0cm}
\includegraphics[width=70mm]{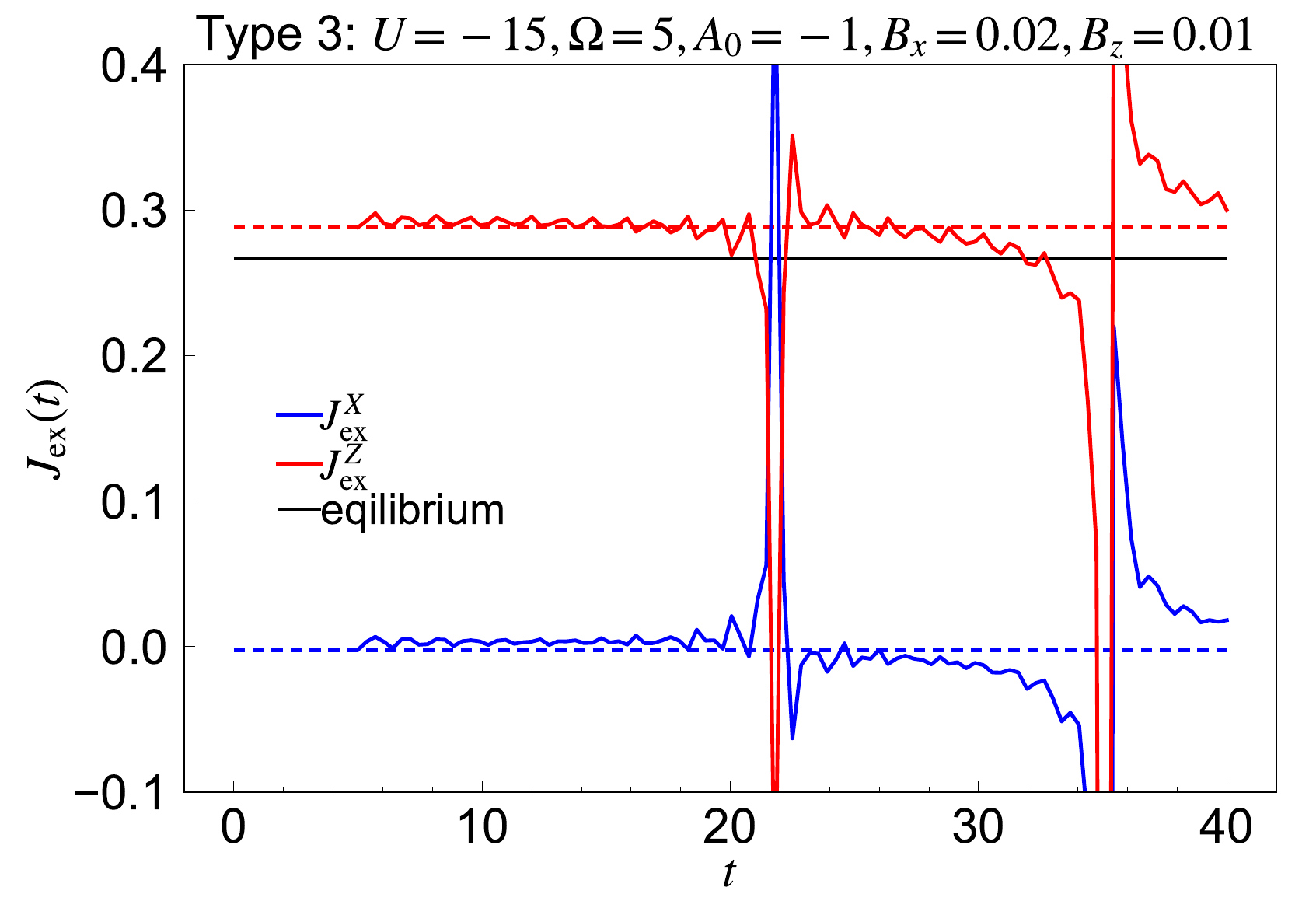} 
  \caption{Time evolution of the exchange coupling evaluated from Eq.~\eqref{eq:eval_J_XXZ} for $U=-15,\Omega=5,A_0=-1,B_x=0.02$ and $B_z=0.01$.
  Here the type 3 excitation with the optimal condition for $l_0=-3,n_1=3$ and $n_2=2$ is used.
 The dashed lines indicate $J_{\eta,XY}^{\rm (HE)}$ and $J_{\eta,Z}^{\rm (HE)}$ (Eq.~\eqref{eq:J_HE_2}).}
  \label{fig:J_ex_attractive}
\end{figure}
%%%%%%%%%%%%%%%%%%%%%%%%%%%%%%%%%%%%%%%%%%%%
As in the case of the AF phase for $U>0$, one can measure the modified exchange couplings of the pseudo-spins for $U<0$  
by comparing the DMFT results with the MF dynamics of the pseudo-spins: $\hH_{\rm \eta-spin} (t)=   J_{\eta,XY}(t)\sum_{(ij)}(\heta^x_i\heta^x_j +  \heta^y_i\heta^y_j) +  J_{\eta,Z} (t)\sum_{(ij)} \heta^z_i\heta^z_j + B_x \sum_i (-)^i  \eta^x_{i} + B_z \sum_i (-)^i \eta^z_{i}$. Physically, the $B_x$ term corresponds to a homogeneous pair potential (remember the $(-)^i$ factor in the definition of $\heta^x_i$), which favors the development of the real SC order parameter.
The $B_z$ term corresponds to a staggered potential, which favors the development of charge order. As in the case of $U>0$, one can consider the MF dynamics of this system,
which yields
\begin{subequations} \label{eq:eval_J_XXZ}
\eqq{
Z J_X & = \frac{1}{2}\Bigl [\frac{B_x}{\eta_x} +  \frac{\dot{\eta}_x} {\eta_y\eta_z} - \frac{\dot{\eta}_y} {\eta_x\eta_z}    \Bigl], \\
Z J_Z & =  \frac{1}{2}\Bigl [\frac{2 B_z}{\eta_z}-\frac{B_x}{\eta_x} +  \frac{\dot{\eta}_x} {\eta_y\eta_z} + \frac{\dot{\eta}_y} {\eta_x\eta_z}    \Bigl].
}
\end{subequations}
In Fig.~\ref{fig:J_ex_attractive},  we show the exchange couplings estimated from these equations for the type 3 protocol with optimal parameters for $U=-15$, $\Omega=5$ and $A_0=-1$.
One can see that the measured $J_{X}$ and $J_{Z}$ well match the predictions $J_{\rm \eta,X}^{\rm (HE)}(=J_{\rm \eta,X}^{\rm (P)})$ and $J_{\rm \eta,Z}^{\rm (HE)}(=J_{\rm \eta,Z}^{\rm (P)})$.
Still, at some times, strong deviations from the predicted values are observed, because $\eta^y$ becomes small for these times.
We note that with other excitation protocols the deviation from the expected values are severe, which can again be attributed to the effect of photo-doping.

\section{Conclusions}\label{sec:conclude}
In this paper, we showed that heating, which is a typically undesired side of the effect Floquet engineering, can be significantly suppressed using multi-color excitation protocols and interference between different excitation pathways in strongly correlated system.
We focused on the one-band Hubbard model and considered sub-gap but strong electric-field excitations with frequency $\Omega$ as the main drive.
As auxiliary excitations, we discussed additional electric-field excitations and/or hopping modulations with frequencies corresponding to higher harmonics of $\Omega$.
Using nonequilibrium DMFT, which is reliable for systems in high spatial dimensions, we studied how the multi-color excitation protocols suppress heating.
We showed that the effective Floquet Hamiltonian in the rotating frame serves as a useful guide in determining the conditions for the efficient suppression, focusing on $3\Omega$- and $2\Omega$-absorption processes.
In practice, an efficient suppression can be realized by suppressing the d-h creation terms in the Floquet Hamiltonian order by order.
We also measured the evolution of the exchange couplings in the driven systems and demonstrated that the suppression of heating removes potentially competing effects associated with photo-doping. Multi-color driving protocols thus allow to more clearly observe the modifications of the low energy physics resulting from virtual excitations, and thus the desired Floquet engineered properties. 

We expect that the physics discussed in this work can be directly observed in cold-atom experiments.~\cite{Kilian2019PRL,Essliinger2021PRX}
An important question is how well the multi-color excitations suppress heating in real materials.
The single band Hubbard model with a large Coulomb interaction can be realized, for example, in  alkali-metal loaded zeolites~\cite{Arita2003PRB} and alkali-cluster-loaded sodalites,\cite{Nakamura2009PRB}
where our results may be directly applied.
For general strongly correlated systems, further analyses of models with multiple orbitals and different lattice structures are required.
Other intriguing issues which may be addressed in the future are the role of the coherence of the excited states and the dimensionality of the system.
The present study implies that the interference between the excitation processes represented by different terms in the effective Hamiltonian
turns out to be not significant in high dimensional systems described by DMFT. Still, one can expect that such interference effects become important when there are only
a few potential final states and the excitations occur before the coherence between the ground state and the excited state is lost.
In Mott insulators in dimensions higher than one, due to the spin-charge coupling, this coherence time is typically short.\cite{Murakami2022PRL}
On the other hand, in lower dimensional systems, in particular in one dimension, the spin-charge coupling become less important and the coherence of the excited states lasts longer.\cite{Murakami2021PRB, Ishihara2022PRR} 
In such a situation, the interference effects originating from the different terms in the Floquet Hamiltonian can be important.

\begin{acknowledgments}
This work is supported by Grant-in-Aid for Scientific Research from JSPS, KAKENHI Grant Nos. JP20K14412 (Y.M.), JP21H05017 (Y.M.), JP19H05825 (R.A.), JST CREST Grant No. JPMJCR1901 (Y.M.), SNSF an Ambizione Grant No. PZ00P2\_193527(M.S.), and ERC Consolidator Grant No.~724103 (P.W.). 
\end{acknowledgments}

\appendix

\section{Results for the two-dimensional square lattice} \label{sec:2d}
In this section, we show supplementary DMFT results for the Hubbard model on the 2d square lattice.
Compared to the case of the Bethe lattice, simulations for the 2d square lattice require more computational resources, since one needs to deal with 
momentum-dependent Green's functions. Thus, it is more difficult to make systematic analyses as we did for the Bethe lattice in the main text.
Still, by comparing some cases, we confirm that the 2d square lattice and the Bethe lattice yield qualitatively the same results.

\begin{figure}[t]
  \centering
    \hspace{-0.cm}
    \vspace{0.0cm}
\includegraphics[width=87mm]{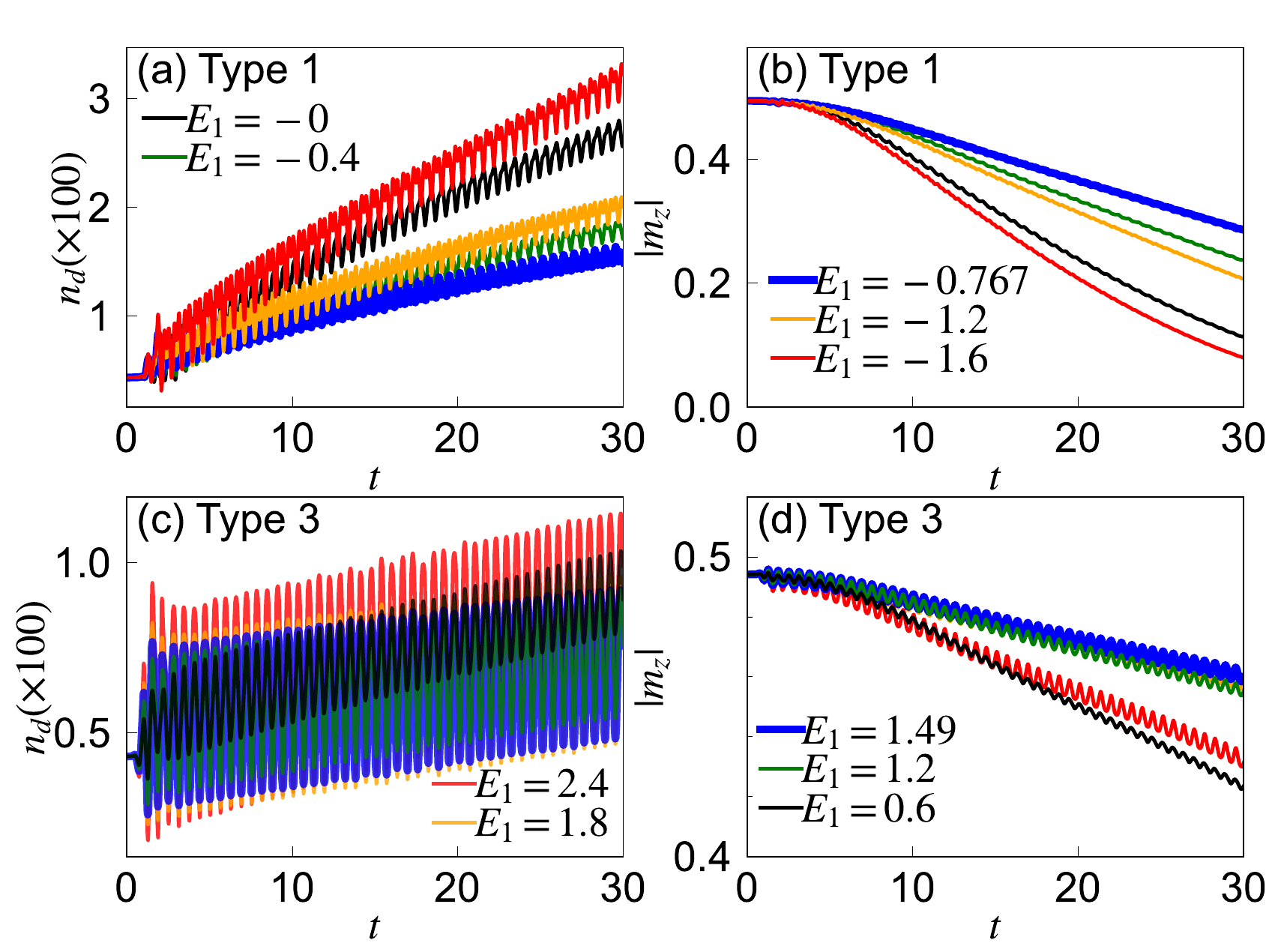} 
  \caption{(a)(c) Evolution of the density of doublons $n_d$ and  (b)(d) the staggered magnetization $m_z$ under periodic excitations with $\Omega=U/3$.
  Here we consider the system on the two-dimensional square lattice and set $v_0=0.5$ and $U=15$.
    The initial temperature is $T=0.02$, so that the system is initially in the AF phase, with the spins pointing along the $z$ direction. The parameters of the main drive are $\Omega=5$ and $A_0=-1$ ($E_0=5$).
  In panels (a) and (b), which share the labels, we use the type 1 protocol with $n_1=3$ and the indicated values of the auxiliary electric field. In (c) and (d), which also share the labels, we use the type 3 protocol with $n_1=3$ and $n_2=2$. The values of the auxiliary electric field and  hopping modulations are chosen such that $\mathcal{B}^{(0)}_{\bf e}$ is zero, i.e. $(E_1,\delta_v)=(0.6,-0.0784)$, $(1.2,-0.112)$, $(1.49,-0.128)$, $(1.8,-0.145)$, $(2.4,$ $-0.177)$. In the figure, only the values of the auxiliary electric field are shown. The thick blue lines correspond to the optimal parameters for each protocol as predicted by the Floquet Hamiltonian. }
  \label{fig:3Ome_each_2dSQ}
\end{figure}

Here we set the hopping parameter to $v_0=0.5$ so that the bandwidth of the free system ($U=0$) is $4$ as in the case of the Bethe lattice.
We consider the field applied to the diagonal of the square lattice. 
In Fig.~\ref{fig:3Ome_each_2dSQ}, we show the evolution of the double occupation $n_d(t)$ and the AF order parameter $m_z(t)$ for $U=15$, $\Omega =5$ and $A_0 = -1$.
These results can be directly compared with those in Fig.~\ref{fig:U_3Ome_tevo} for the Bethe lattice, and they show the qualitatively same behavior.
In particular, we observe the strongest suppression of heating for the optimal condition predicted by the Floquet Hamiltonian.

 %%%%%%%%%%%%%%%%%%%%%%%%%%%%%%%%%%%%%%%%%%%%%
%%%%%%%%%%%%%%%%%%%%%%%%%%%%%%%%%%%%%%%%%%%%

\section{Expressions for $\hH_\text{eff}^{(2)}$ }\label{sec:H_eff}
In this section, we present the explicit expressions for  $\hH_\text{eff,2}^{(2)}$ and  $\hH_\text{eff,3}^{(2)}$.
For $\hH_\text{eff,2}^{(2)}$, we need to evaluate $[\hat{g}_{ij\sigma},\hat{h}^\dagger_{i'j'\sigma'}]$, which can be expressed as follows:
\begin{enumerate}
\item $i'=i,\; j'=j$ : 0 
\item $i'=j,\; j'=i$ : 0 
\item $i'=i,\; j'\neq j$ 
\begin{enumerate}
\item $\sigma'=\sigma$ : 0
\item $\sigma'=\bar{\sigma}$ : $-c^\dagger_{i\sigma} c_{j\sigma} c^\dagger_{i\bar{\sigma}} c_{j'\bar{\sigma}} \bar{n}_{j\bar{\sigma}} \bar{n}_{j'\sigma}$
\end{enumerate}
\item $i'\neq i,\; j' = j$ 
\begin{enumerate}
\item $\sigma'=\sigma$ : 0
\item $\sigma'=\bar{\sigma}$ : $-c^\dagger_{i\sigma} c_{j\sigma} c^\dagger_{i'\bar{\sigma}} c_{j\bar{\sigma}} n_{i\bar{\sigma}} n_{i'\sigma}$
\end{enumerate}
\item $i'=j,\; j' \neq i$ 
\begin{enumerate}
\item $\sigma'=\sigma$ : $c^\dagger_{i\sigma} c_{j'\sigma} n_{i\bar{\sigma}} n_{j\bar{\sigma}} \bar{n}_{j'\bar{\sigma}}$
\item $\sigma'=\bar{\sigma}$ : $c^\dagger_{j\bar{\sigma}} c_{j'\bar{\sigma}} c^\dagger_{i \sigma} c_{j \sigma} \bar{n}_{j' \sigma} n_{i \bar{\sigma}}$
\end{enumerate}
\item $i'\neq j,\; j' = i$ 
\begin{enumerate}
\item $\sigma'=\sigma$ :  $-c^\dagger_{i'\sigma} c_{j\sigma} \bar{n}_{i\bar{\sigma}} \bar{n}_{j\bar{\sigma}} n_{i'\bar{\sigma}}$
\item $\sigma'=\bar{\sigma}$ :  $c^\dagger_{i'\bar{\sigma}} c_{i\bar{\sigma}} c^\dagger_{i \sigma} c_{j \sigma} n_{i' \sigma} \bar{n}_{j \bar{\sigma}}$
\end{enumerate}
\end{enumerate}
In the other cases, the commutator becomes zero. $[\hat{g}_{ij\sigma},\hat{h}_{i'j'\sigma'}]$ can be obtained by considering the Hermitian conjugate.

For $\hH_\text{eff,3}^{(2)}$, we need to evaluate $[\hat{h}^\dagger_{ij\sigma},\hat{h}_{i'j'\sigma'}]$, which can be expressed as follows:

\begin{enumerate}
\item $i'=i,\; j'=j$
\begin{enumerate}
\item $\sigma'=\sigma$ : $n_{i\bar{\sigma}} \bar{n}_{j\bar{\sigma}} ( n_{i\sigma}- n_{j\sigma})$\\
This term corresponds to the exchange couplings of the Z component in the spin Hamiltonian~\eqref{eq:H_spin} and the pseudo-spin Hamiltonian~\eqref{eq:H_eta}.
\item $\sigma'=\bar{\sigma}$ : $-c^\dagger_{j\bar{\sigma}} c_{i\bar{\sigma}} c^\dagger_{i\sigma} c_{j\sigma}$\\
This term corresponds to  the exchange coupling of the XY component of the spin Hamiltonian~\eqref{eq:H_spin}.
\end{enumerate}
\item $i'=j,\; j'=i$
\begin{enumerate}
\item $\sigma'=\sigma$ : 0
\item $\sigma'=\bar{\sigma}$ : $c^\dagger_{i\sigma} c_{j\sigma} c^\dagger_{i\bar{\sigma}} c_{j\bar{\sigma}}$\\
This term corresponds to the exchange coupling of the XY component of the pseudo-spin Hamiltonian~\eqref{eq:H_eta}.
\end{enumerate}
\item $i'=i,\; j'\neq j$
\begin{enumerate}
\item $\sigma'=\sigma$ : $-c^\dagger_{j'\sigma} c_{j\sigma} n_{i\bar{\sigma}} \bar{n}_{j\bar{\sigma}} \bar{n}_{j'\bar{\sigma}}$
\item $\sigma'=\bar{\sigma}$ : $-c^\dagger_{j'\bar{\sigma}}c_{i\bar{\sigma}} c^\dagger_{i\sigma} c_{j\sigma} \bar{n}_{j'\sigma} \bar{n}_{j\bar{\sigma}}$
\end{enumerate}
\item $i'\neq i,\; j'=j$
\begin{enumerate}
\item $\sigma'=\sigma$ : $c^\dagger_{i\sigma} c_{i'\sigma} n_{i\bar{\sigma}} \bar{n}_{j\bar{\sigma}} n_{i'\bar{\sigma}}$
\item $\sigma'=\bar{\sigma}$: $-c^\dagger_{j\bar{\sigma}} c_{i'\bar{\sigma}} c^\dagger_{i\sigma} c_{j\sigma} n_{i'\sigma} n_{i\bar{\sigma}}$
\end{enumerate}
\item $i'=j,\; j'\neq i$
\begin{enumerate}
\item $\sigma'=\sigma$ : $0$
\item $\sigma'=\bar{\sigma}$ : $c^\dagger_{i\sigma} c_{j\sigma} c^\dagger_{j'\bar{\sigma}} c_{j\bar{\sigma}} n_{i\bar{\sigma}} \bar{n}_{j'\sigma}$
\end{enumerate}
\item $i'\neq j,\; j'=i$
\begin{enumerate}
\item $\sigma'=\sigma$ : $0$
\item $\sigma'=\bar{\sigma}$ : $c^\dagger_{i\sigma} c_{j\sigma} c^\dagger_{i\bar{\sigma}} c_{i'\bar{\sigma}} \bar{n}_{j\bar{\sigma}} n_{i'\sigma}$
\end{enumerate}
\end{enumerate}

 %%%%%%%%%%%%%%%%%%%%%%%%%%%%%%%%%%%%%%%%%%%%%
 \begin{figure}[b]
  \centering
    \hspace{-0.cm}
    \vspace{0.0cm}
\includegraphics[width=80mm]{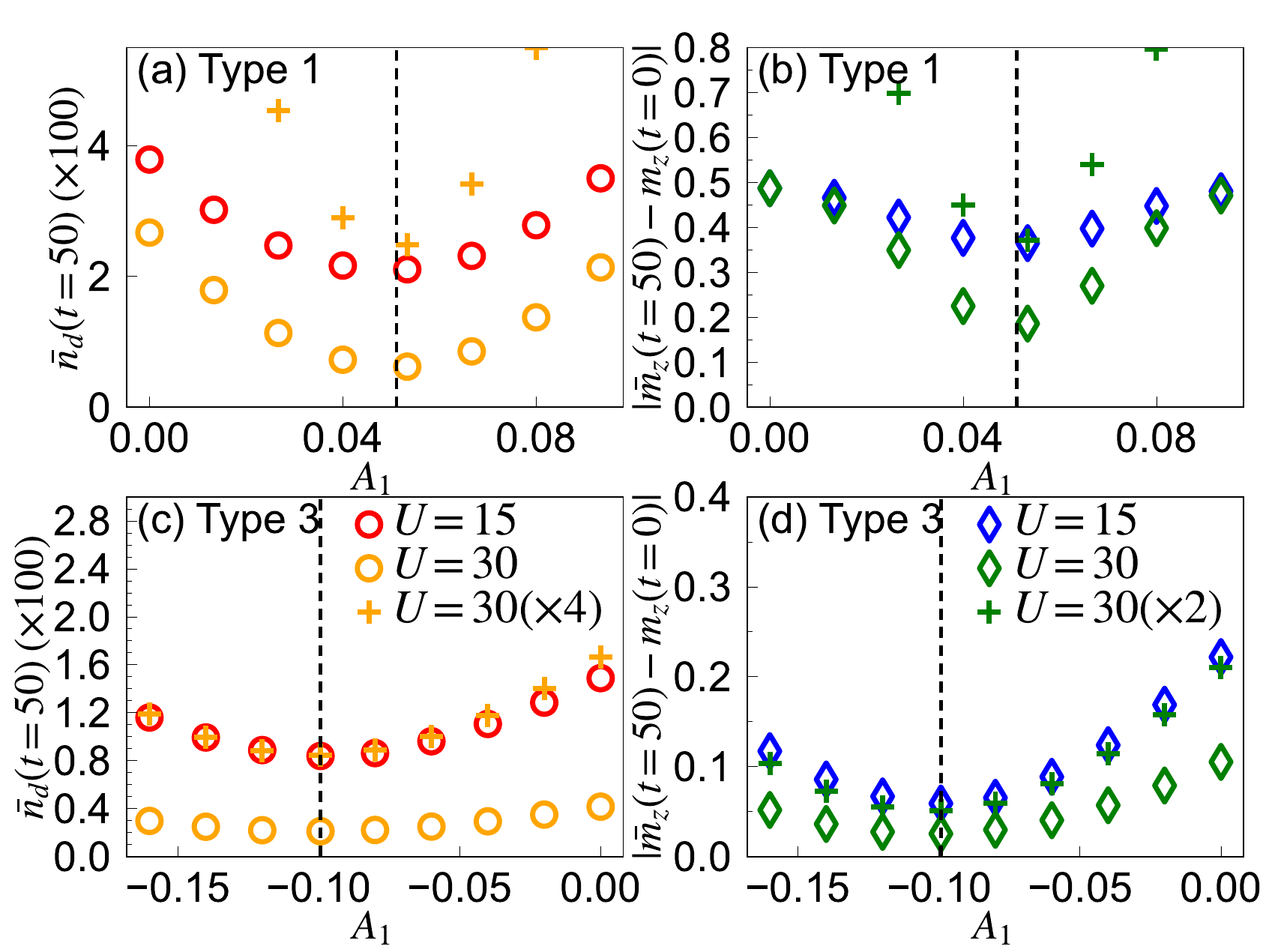} 
  \caption{Comparison of the doublon number and the staggered magnetization for $U=15$ and $U=30$. Both are averaged over a time interval of length $\frac{2\pi}{\Omega}$ around $t=50$.
  The parameters of the main electric field excitation are $\Omega=U/3$ and $A_0=-1$.  We set $l_0=3,n_1=3$ and $n_2=2$. The dashed lines indicate the optimal conditions for each protocol. }
  \label{fig:U_dep}
\end{figure}
%%%%%%%%%%%%%%%%%%%%%%%%%%%%%%%%%%%%%%%%%%%%

\section{Scaling with $U$} \label{sec:U_scale}

Here, we discuss how the behavior of relevant physical quantities scales with $U$.
In Fig.~\ref{fig:U_dep}, we compare systems with $U=15$ and $U=30(=15\times2)$ and the type 1 and type 3 protocols.
For $U=15$, the excitation conditions are the same as those in Figs. \ref{fig:U_3Ome_tevo} and \ref{fig:U_3Ome_tevo_summary}.
For $U=30$, the excitation frequencies are rescaled by factor 2, keeping the values of $A_0$, $A_1$ and $\delta_v$.
We note that, as in the case of $U=15$, for $U=30$, the heating is efficiently suppressed around the optimal conditions predicted by the Floquet Hamiltonian.
In the type 1 protocol, the doublon number scales as $\mathcal{O}(1/U^2)$ in the optimal condition.
On the other hand, away from the optimal condition, the difference from the optimal case does not change with $U$.
The reduction of the order parameter from the equilibrium value scales as $\mathcal{O}(1/U)$ in the optimal condition.
Away from the optimal condition, the difference from the optimal case is larger for $U=30$.
In the type 3 protocol, the doublon number scales as $\mathcal{O}(1/U^2)$, while the reduction of the order parameter scales as $\mathcal{O}(1/U)$.

We can understand the above behavior as follows.
First, the equilibrium doublon number scales as $\mathcal{O}(1/U^2)$, and the energy scale of the AF order scales as $\mathcal{O}(1/U)$. 
In the type 1 protocol with the optimal condition, the d-h creation term in the Floquet Hamiltonian scales as $\mathcal{O}(1/U)$ $(= \mathcal{O}(1/\Omega))$,
since the leading order terms are canceled. Away from the optimal condition, the d-h creation term scales as  $\mathcal{O}(1)$.
In the type 3 protocol, the d-h creation term scales as $\mathcal{O}(1/U)$  $(= \mathcal{O}(1/\Omega))$. 
Note that in this protocol, the optimal condition is chosen to minimize the d-h  terms at the order of $1/\Omega$, but it never makes them zero. 
If we apply the Fermi Golden rule for the d-h creation process, the d-h creation rate scales with the square of the coefficients of the d-h creation terms.
The heating rate should also scale similarly. 
Moreover, the magnitude of the order reduction is determined by the ratio between the heating rate and the energy scale of the order.
Thus, for example, in the type 3 protocol, the doublon number is expected to scale as $\mathcal{O}(1/U^2)$.
On the other hand, given that the energy scale of the AF order is $\mathcal{O}(1/U)$, the reduction of the magnetization should behave as  $\mathcal{O}(1/U)$.
This consistently explains the numerical results. The behavior for the type 1 protocol can be explained analogously.

\section{Spectral function under excitations} \label{sec:Aw_floquet}
\begin{figure}[t]
  \centering
    \hspace{-0.cm}
    \vspace{0.0cm}
\includegraphics[width=70mm]{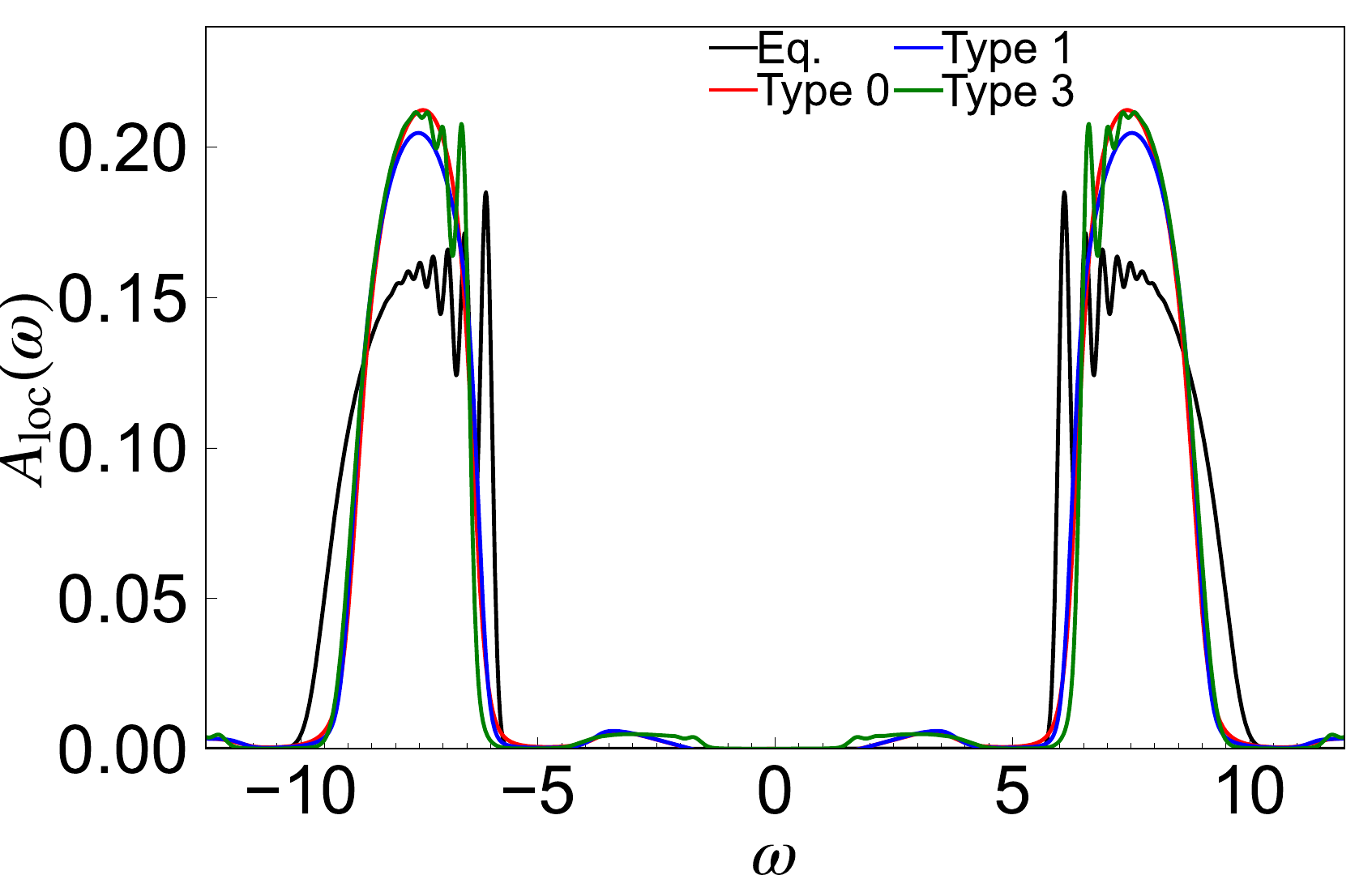} 
  \caption{Local single-particle spectral functions for $U=15$ and $T=0.02$ for indicated cases. The parameters of the  main electric field excitation are $\Omega=5$ and $A_0=-1$.
  For auxiliary fields, we set $n_1=3$ and $n_2=2$  and use the optimal conditions determined from the Floquet Hamiltonian. }
  \label{fig:Aw_reply}
\end{figure}

In Fig.~\ref{fig:Aw_reply}, we compare the single-particle spectra $A_{\rm loc}(\omega)$ evaluated for various cases.
Here, the spectrum is defined as 
$A^R_{\rm loc}(\omega)\equiv - \frac{1}{\pi}\text{Im} \int dt e^{i\omega (t_0-t)} G^R_{\rm loc}(t_0,t),$
where $G^R_{\rm loc}(t_0,t)$ is the retarded part of the local Green's function averaged over spins. We set $t_0\simeq 60$.
As we explained in Sec.~\ref{sec:equilibrium}, in equilibrium, there emerge peaks corresponding to the formation of spin polarons in the AF phase.
Under the electric field, the width of the Hubbard bands is reduced due to the dynamical localization effect.
The renormalized bandwidth is almost the same among the type 0, type 1 and type 3 protocols, which indicates that the effect of the auxiliary fields is minor here.
Reflecting that the AF order is  robust against the excitation in the type 3 protocol, one can identify the spin polarons in the spectrum.
On the other hand, in the other two protocols, the AF order is suppressed, and no spin polaron structures can be identified.

 %%%%%%%%%%%%%%%%%%%%%%%%%%%%%%%%%%%%%%%%%%%%%
%%%%%%%%%%%%%%%%%%%%%%%%%%%%%%%%%%%%%%%%%%%%

\bibliography{HHG_Ref}

\end{document}